\pdfoutput=1                                                                  %aa.dem
\documentclass[twocolumn]{aa} % for a paper on 2 columns  
\usepackage{amsmath} % for align
\usepackage{graphicx}
\usepackage{xcolor}
\usepackage{natbib}
\usepackage[varg]{txfonts}
\usepackage{mathabx}
\usepackage{subcaption}

\usepackage{comment}
\usepackage{ulem}
\usepackage{lineno}

\usepackage{booktabs,caption}
\usepackage[flushleft]{threeparttable}

\bibpunct{(}{)}{;}{a}{}{,} % to follow the A&A style
\usepackage[breaklinks, colorlinks, citecolor=blue, linkcolor=blue]{hyperref}
\makeatletter
\def\instrefs#1{{\def\scsep{\def\scsep{,}}\@for\w:=#1\do{\scsep\ref{inst:\w}}}}
\renewcommand{\inst}[1]{\unskip$^{\instrefs{#1}}$}
\graphicspath{{./}{./figures/}}

\newcommand{\gj}{GJ~486}
\newcommand{\gjb}{GJ~486b}
\def\degs{\ifmmode ^{\circ}\else$^{\circ}$\fi}

\newcommand{\Msun}{\ensuremath{M_{\sun}}}

\newcommand{\Rp}{\ensuremath{R_{\text{pl}}}}
\newcommand{\Bp}{\ensuremath{B_{\text{pl}}}}
\newcommand{\Rmp}{\ensuremath{R_{\text{mp}}}}
\newcommand{\kmp}{\ensuremath{k_{\text{mp}}}}
\newcommand{\vrel}{\ensuremath{v_{\text{rel}}}}

\newcommand{\nsw}{\ensuremath{n_{\text{sw}}}}
\newcommand{\pdynsw}{\ensuremath{p_{\text{dyn,sw}}}}
\newcommand{\pthsw}{\ensuremath{p_{\text{th,sw}}}}
\newcommand{\Tc}{\ensuremath{T_{\text{c}}}}
\newcommand{\Bsw}{\ensuremath{B_{\text{sw}}}}
\newcommand{\Mdotstar}{\ensuremath{\dot{M}_\star}}
\newcommand{\Mdotsun}{\ensuremath{\dot{M}_\sun}}

\begin{document}

\title{Searching for star-planet interactions in GJ~486 at radio wavelengths with the uGMRT}
\titlerunning{Searching for star-planet interactions in GJ~486 at radio wavelengths with the uGMRT}
\authorrunning{Pe\~na-Mo\~nino et al.}

 \author{L.~Pe\~na-Mo\~nino{\inst{iaa}} \and
         M.~P\'erez-Torres{\inst{iaa,euc}}  \and
         D.~Kansabanik{\inst{ucar}} \and
         G.~Blázquez-Calero{\inst{iaa}} \and
         R.\,D.~Kavanagh{\inst{ASTRON,Anton}}  \and 
         J.\,F.~Gómez{\inst{iaa}} \and
         J.~Moldón {\inst{iaa}} \and
         A.~Alberdi{\inst{iaa}} \and
         P.\,J.~Amado{\inst{iaa}} \and
         G.~Anglada{\inst{iaa}} \and
         J.\,A.~Caballero{\inst{CAB}} \and
         A.~Mohan{\inst{godd}}  \and
         P.~Leto{\inst{INAF}} \and
         M.~Narang {\inst{asiaa}} \and
         M.~Osorio{\inst{iaa}} \and
         D.~Revilla{\inst{iaa}} \and
         C.~Trigilio{\inst{INAF}} 
        }

 \institute{
 IAA-CSIC, Instituto de Astrof\'isica de Andaluc\'ia, 
Glorieta de la Astronom\'ia s/n, 18008, Granada, Spain \label{inst:iaa} 
    \\
    \email{\href{mailto:lpm@iaa.es}{lpm@iaa.es}} 
   \and
   School of Sciences, European University Cyprus, Diogenes street, Engomi, 1516 Nicosia, Cyprus
   \label{inst:euc}
   \and 
   Cooperative Programs for the Advancement of Earth System Science, University Corporation for Atmospheric Research, Boulder, CO, USA \label{inst:ucar}
   \and
ASTRON, Netherlands Institute for Radio Astronomy, Oude Hoogeveensedijk 4, Dwingeloo, 7991 PD, The Netherlands \label{inst:ASTRON}   
   \and
   Anton Pannekoek Institute for Astronomy, University of Amsterdam, 1090 GE Amsterdam, the Netherlands\label{inst:Anton}
   \and
   Academia Sinica Institute of Astronomy \& Astrophysics, Taipei 10617, Taiwan \label{inst:asiaa}
   \and
    Goddard Space Flight Center, Greenbelt, Maryland, USA \label{inst:godd}
    \and
    INAF-Osservatorio Astrofisico di Catania, Via Santa Sofia 78, 95123 Catania, Italy \label{inst:INAF}
    \and
    Centro de Astrobiología (CSIC-INTA), Campus ESAC, Camino Bajo del Castillo s/n, 28692 Villanueva de la Cañada, Madrid,Spain \label{inst:CAB}
   \\
}

\date{Received 8 August 2024; accepted 21 November 2024}

\abstract{}
{We search for radio emission from star-planet interactions in the M-dwarf system \gj, which hosts an Earth-like planet. }
{We observed the GJ~486 system with the upgraded Giant Metrewave Radio Telescope (uGMRT) from 550 to 750 MHz in nine different epochs, between October 2021 and February 2022, covering almost all orbital phases
of \gjb\ from different orbital cycles. We obtained radio images and dynamic spectra of the total and circularly polarized intensity for each individual epoch.} 
{We do not detect any quiescent radio emission in any epoch above 3$\sigma$.
Similarly, we do not detect any bursty emission in our dynamic spectra.
{
While we cannot completely rule out that the absence of a radio detection is due to time variability of the radio emission, or to the maximum electron-cyclotron maser emission being below our observing range, this seems unlikely. We discuss two possible scenarios: an intrinsic dim radio signal, or alternatively, that the anisotropic beamed emission pointed away from the observer. 
If the non-detection of radio emission from star-planet interaction in GJ~486 is due to an intrinsically dim signal, this implies that, independently of whether the planet is magnetized or not, the mass-loss rate is small (\Mdotstar $\lesssim$ 0.3 \Mdotsun) and that, concomitantly, the efficiency of the conversion of Poynting flux into radio emission must be low ($\beta \lesssim 10^{-3}$).
Free-free absorption effects are negligible, given the high value of the coronal temperature. 
Finally, if the anisotropic beaming pointed away from us, this would imply that GJ~486 has very low values of its magnetic obliquity and inclination. 
}
}
{}


\maketitle
%
%-------------------------------------------------------------------
\section{Introduction} \label{sec:intro}
 
M dwarfs, also known as ``red dwarfs'', are the most common stars in the solar
neighbourhood (e.g., \citealt{Reyle2021}). They are expected to
preferentially host close-in rocky planets \citep{Sabotta2021, Ribas2023}.  The discovery of
Proxima b, an Earth-sized, rocky planet within the habitability zone of its star
Proxima Centauri \citep{Anglada-Escude2016},  boosted the interest of the astronomical
community, and the study of exoplanets and their potential habitability has blossomed in
the last decade, thanks to space-based missions (e.g., Kepler, TESS), as well as to the
exploitation of data taken with ground-based telescopes.

The discovery of the majority of exoplanets around M dwarfs, including rocky ones, has
been done using radial velocity and transit measurements, while their discovery in the
radio has been elusive despite decades of relentless efforts.  The detection of auroral
radio emission from an exoplanet would confirm the existence of such an exoplanet, and
would inform us of its intrinsic magnetic field.  The latter is a piece of information
that cannot be provided by other means, which makes radio observations unique.  
Unfortunately, Earth and super-Earth exoplanets with plausible magnetic field strengths
of up to a few Gauss would result in auroral radio emission from such an exoplanet at
frequencies below the Earth's ionospheric cutoff of $\sim$10 MHz, therefore making this
radio emission undetectable from Earth.  However, if an exoplanet is close enough to its
host star that it is in the sub-Alfvénic regime, i.e., when the plasma speed relative to
the exoplanet, $v$, is smaller that the Alfvén speed at the planet position, $v_A$, then
energy and momentum can be transported upstream back to the star by Alfvén waves.
Jupiter’s interaction with its Galilean satellites is a well-known example of
sub-Alfvénic interaction, which results in copious auroral radio emission
\citep{Zarka2007}.  The mechanism responsible for this emission is the electron-cyclotron maser (ECM) instability, which is a coherent mechanism yielding strong, variable, highly-polarized emission. The typical energies of the electrons involved in auroral radio emission are around 1 keV
\citep{Wu1979}, and for the case of the Io-Jupiter decameter emission, the range of energies can go up to about 20 keV (e.g. \citealt{Lamy2022}). 
In the case of magnetic star-planet  interaction (SPI), the radio emission
arises  from the upper layers of the atmosphere of the host star, induced by the
exoplanet moving through the stellar magnetosphere, and thus the relevant magnetic field
is that of the star, $B_\star$, not the exoplanet magnetic field. This makes a huge
difference in the frequency where to search for this radio emission.

M dwarfs are excellent candidates to detect star-planet interaction at radio
wavelenghts, since they have surface magnetic fields of a few hundred Gauss, or even
kGauss \citep{Shulyak2019,Kochukhov2021,Reiners2022}, so the fundamental frequency of the cyclotron emission
falls into the detection capabilities of existing radio interferometers.  Sub-Alfvénic
interaction is expected to yield detectable periodic radio emission via the
ECM mechanism,  as long as the Alfvén wave connects
back to the star (e.g.  \citealt{Zarka2007, Saur2013, Turnpenney2018, Vedantham2020,
PerezTorres2021}). The confirmation of this radio emission would provide a completely new method of exoplanet detection. Namely, one would expect to detect strong, highly circularly
polarized radio emission at an observing frequency close to, or below, the maximum
cyclotron frequency, and showing a periodic signal that correlates with the orbital
period of the planet, $P_{\rm orb}$, or the synodic period, $P_{\rm syn}^{-1} = P_{\rm
rot}^{-1} - P_{\rm orb}^{-1}$, where $P_{\rm rot}$ is the rotation period of the host
star, although we note that the geometry of the system  can result in a more complex
relation. 

In recent years, there have been several claims of exoplanet induced radio emission
(GJ~1151b - \citealt{Vedantham2020}; Proxima b - \citealt{PerezTorres2021}; YZ Cet b -
\citealt{Pineda2023,Trigilio2023}), but a solid confirmation is still lacking for those
cases. For example, \citet{Narang2024} recently reported non-detections of GJ~1151 with the uGMRT at 150, 218, and 400 MHz.
In this paper, we present the results from the longest monitoring radio campaign
on the star-exoplanet M dwarf system GJ~486 - GJ~486b, using the upgraded Giant
Metrewave Radio Telescope, uGMRT \citep{Gupta2017}. Our overall aim is to characterize
the radio emission from this system, with the main goal of detecting radio emission due to magnetic SPI.  We describe the GJ~486 -- GJ~486b system in Section
\ref{sec:target}, and  our observations and data analysis in Section \ref{sec:obs}. We
present our results in Section \ref{sec:results} and discuss them in Section
\ref{sec:discussion}.

\section{The GJ~486 -- GJ~486b system} \label{sec:target}

\gj\ is a cool, nearby M3.5V star 
 (\citealt{Caballero2022}; see also Table \ref{tab:gj486-params}). It rotates at a rate roughly half of the Sun and, given its mass, it is very likely an almost fully convective star.
\citet{Trifonov2021} reported the existence of an exoplanet, \gjb, using radial velocity (RV) measurements from the CARMENES survey \citep{Quirrenbach2016} and TESS photometry \citep{Ricker2015}.
\gjb\ has a radius and a mass slightly larger than the Earth, straddling the boundary between Earth-like planets and super-Earths. However, its bulk density ($\rho \approx 7.0$ g\,cm$^{-3}$) favors a massive terrestrial planet rather than an ocean one \citep{Trifonov2021}. 
Given the proximity of the GJ 486 system (8.1 pc from Earth), and the small separation of the planet to its host star ($\lesssim$ 0.02\,au), GJ 486 showed promise for detecting radio emission arising from star-planet interaction.

\begin{table}[htb]
 \begin{threeparttable}
\caption{Parameters of the GJ~486 -- GJ~486b system}
\label{tab:gj486-params}
\begin{tabular}{ll}
\toprule
Parameter  & Value \\ 
\toprule
\multicolumn{2}{l}{Star (GJ 486)} \\ 
\cline{1-1}
$\alpha$ (J2000)     & 12:47:56.62\\ 
$\delta$ (J2000)      & +09:45:05.0\\ 
$\mu_{\alpha}$ cos $\delta$ [mas\ yr$^{-1}$] & 	-1008.267 $\pm$ 0.040 \\
$\mu_{\delta}$ [mas\ yr$^{-1}$]   &  	-460.034 $\pm$ 0.033 	\\
Spectral Type & M3.5V \\ 
$T_{\rm eff}$ (K)   & 3291 $\pm$ 75 \\
$M_\star$/$M_{\odot}$     & 0.333 $\pm$ 0.019  \\ 
$R_\star$/$R_{\odot}$     & 0.339 $\pm$ 0.015 \\ 
Distance [pc]     & 8.0827 $\pm$ 0.0021  \\ 
$P_{\rm rot}$ [days]   & 49.9 $\pm$ 5.5  \\ 
\toprule
\multicolumn{2}{l}{Planet (GJ 486b)} \\ \cline{1-1}
$P_{\rm orb}$ [days]          & $1.4671205^{+0.0000012}_{-0.0000011}$       \\ 
$a$ [au]         & $0.01713^{+0.00091}_{-0.00098}$             \\ 
$i$ [degrees]   & 88.90$^{+0.69}_{-0.84}$\\
$M_{\rm p}$/$M_{\oplus}$         & $3.00^{+0.13}_{-0.13}$                    \\ 
$R_{\rm p}$/$R_{\oplus}$           & $1.343^{+0.063}_{-0.062}$                   \\

\bottomrule
\end{tabular}
\begin{tablenotes}
    \small 
    \item All values from \citet{Caballero2022}.
\end{tablenotes}
 \end{threeparttable}
\end{table}

\subsection{ECM frequency and magnetic field of GJ~486}\label{subsec:radio}

If  radio emission is produced on a planet via the ECM instability, then the intrinsic planetary magnetic field can be directly measured.
The maximum cyclotron frequency is  given by

\begin{equation} \label{eq:nu-cycl}
 \nu_c [{\rm  MHz}] = s \times 2.8\, B ,
\end{equation}

\noindent where $B$ is the local magnetic field, in Gauss, and $s$ is the
harmonic number of the cyclotron frequency. ECM emission is observed
essentially from the fundamental ($s = 1$)
\citep{Melrose1982}.  
However, the detection of radio emission from star-planet interaction could be
feasible if the planet is in the sub-Alfvénic regime, as mentioned in
Sect.~\ref{sec:intro}. In this case, the relevant local magnetic field in
Eq.~\ref{eq:nu-cycl} is that of the star, $B_\star$. Using
high-resolution spectral measurements with CFHT/ESPaDOnS, \cite{Moutou2017}
determined a total surface magnetic field value of  $B_\star\, f = 1.6\pm$0.3 kG for
GJ~486, where $f$ is the filling factor  (the fraction of the stellar surface
covered with magnetic field).
Two filling factors could be considered here, namely the Stokes I and V filling factors, $f_I$ and $f_V$ respectively \citep{Morin2008}. However, the relevant one is $f_V$, which refers to the fraction of the field that manifests as a large-scale surface magnetic field, as opposed to $f_I$, which also takes into account smaller-scale field that may cancel out at larger scales, and is much more poorly constrained.
From spectropolarimetric observations of a sample of mid M-dwarf stars of spectral types from M3 to M4.5, \citet{Morin2008} found that
$f_V$ is in the range from 0.10 to 0.15, with typical uncertainties of 0.03 in those values. Here, we used $f = 0.15$, which results in an average stellar
magnetic field value of $B_\star = 240 \pm 45$ G. 
With this value of $B_\star$, 
the maximum cyclotron frequency would then be  $\nu_c = 2.8\, B_\star = 672 \pm 126 $ MHz, and $\nu_c = 1344 \pm 252 $ MHz for $s=1$ (fundamental ECM emission)
and  $s=2$ (second harmonic), respectively. 
We note that, since $B_\star = 240 \pm 45$ G is the average value of the magnetic field in the surface of the star, the value near the (magnetic) poles of the star, where the ECM is most likely produced, has a larger value. On the other hand, ECM emission can be generated at some height above the stellar surface, where the local magnetic field would be smaller. We therefore assumed a nominal value of $B_\star = 240$ G, 
and carried out observations with the uGMRT in band 4 (550 – 900 MHz), which covers the relevant frequency range to search for ECM radio emission that could
arise from star-planet interaction in the GJ~486 system.

\subsection{X-rays from GJ~486} 
\label{subsec:x-rays}

Stellar X-ray and radio (quiescent) activity correlate well \citep{Guedel1993}. Hence,  host stars that are strong X-ray emitters would be expected to be also strong radio emitters, which could jeopardize the search for radio emission from star-planet interaction.

GJ~486 was not detected in X-rays with ROSAT (upper limit of $f_x = 5.37\times10^{-14}$ erg s$^{-1}$cm$^{-2}$;  \citealt{Stelzer2013}). However, recent XMM-Newton observations by \citet{Sanz-Forcada2024} resulted in a clear detection of GJ~486 in X-rays, with a flux at Earth of  $f_x = 4.05\times10^{-15}$ erg s$^{-1}$cm$^{-2}$. 
Using the relationship between radio and X-ray luminosities for M dwarf stars from \citet{Guedel1993}, the X-ray luminosity of GJ~486 would result in a corresponding radio flux density of the star of a few $\mu$Jy. Therefore, the expected quiescent radio emission from the star should be negligible.

%%%%%%%%%%%%%%%%% OBSERVATIONS %%%%%%%%%%%%%%%%%%%%
\section{Observations and data processing} \label{sec:obs}

\begin{figure*}[htb]
  \label{fig:orb_coverage}
  \centerline{\includegraphics[width=\linewidth]{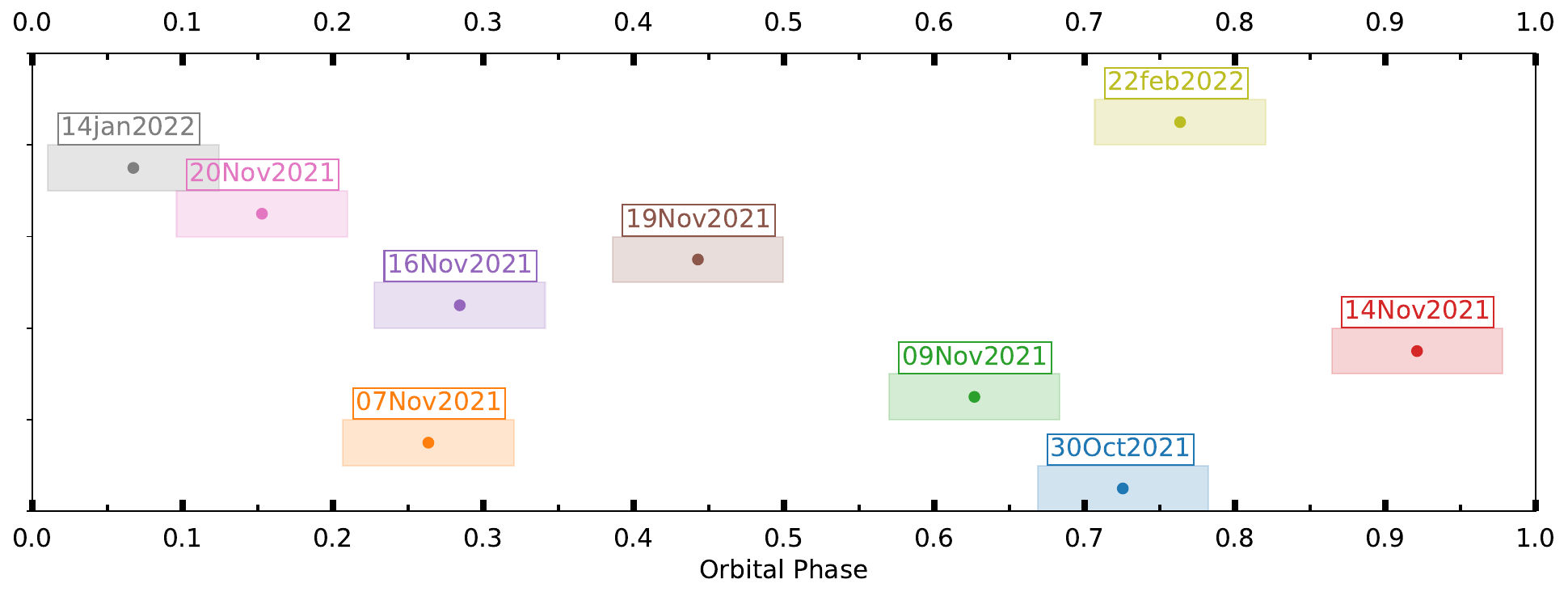}}
  \caption{\label{fig:orb_phase_coverage} \small Coverage of our uGMRT observations of the GJ~486 system, folded to the orbital period of \gjb\ of 1.47 d. Each epoch is shown with a different color. Each point corresponds to the central time of each observing epoch, with the horizontal side of the surrounding rectangle being the time span of the epoch. The vertical axis has no real meaning and the epochs are simply offset so they do not overlap.
}
\end{figure*}

 \noindent We observed  \gj\  with uGMRT on 30 October, 7, 9, 14, 16, 19 and 20 November 2021, and 14 January and 22 February 2022. 
We show in Fig. \ref{fig:orb_phase_coverage} the orbital phase coverage of our observations, which  span almost the entire orbital phase of \gjb, 
and spread out over more than two full rotation periods of the star.
 
We observed \gj\ at a central frequency of 650 MHz, using a bandwidth of 200
MHz and a standard integration time of 10.7 s.  We used 2048 channels in
total, for a channel width of 97.7 kHz.  Given the high proper motion of \gj\ (Table \ref{tab:gj486-params}),
we centered all our
observations at $\alpha$(J2000.0) = 12h 47m 56.00s and 
$\delta$(J2000.0) = $09^\circ$ $45'$ $05.00''$. 
Each observing epoch lasted
typically three to four hours, and we used a duty cycle of about 35 minutes,
with 30 minutes on our target  and 5 minutes on the quasar 3C~286, which we
used as bandpass, flux and phase calibrator. The angular distance between this calibrator and \gj\ is of $\approx$ 23\degs .

We performed all calibration steps using the Common Astronomy
Software Application (CASA; \citealt{McMullin2007,CASATeam2022}), version
casa-6.2.1-7-pipeline-2021.2.0.128. 
First, we used task \texttt{flagdata} to flag edge frequency channels of the observing band as well as frequency channels affected by strong radio frequency interference (RFI). Followed by this, we flagged dead antennas as well as 10 seconds at the start and end of each scan. We did this initial flagging on both 3C 286 and out target field, GJ 486. 
Afterwards, 
we performed an initial delay, bandpass and gain calibration, using 3C~286. First, we run the task \texttt{gaincal} to perform the delay calibration, using the entire bandwidth. 
We then carried out a time-dependent gain calibration using again \texttt{gaincal}, but just over a small part of RFI-free band to minimize any frequency dependence. After applying the delay and time dependent gain solution on-the-fly, we estimated frequency dependent gain solutions using the task \texttt{bandpass}.
After these initial calibration steps, we did automated flagging using  the \texttt{tfcrop} and \texttt{rflag} tasks on the residual visibilities to remove any low-level RFI, and repeated the previous calibration steps. After three rounds of calibration and flagging, we found that the residual visibilities of 3C 286 looked clean. We finally applied the calibration solutions of the final round to our target, GJ~486.
We then performed one round of automated flagging using \texttt{tfcrop} and \texttt{rflag} on the corrected visibilities of the target field. Since the central square antennas of the uGMRT 
have strong correlated RFIs, we performed this automated flagging step on the central square baselines and other baselines, separately.

We performed the imaging and self-calibration of our calibrated data with a modified version of the \texttt{CAPTURE} (CAsa Pipeline-cum-Toolkit for Upgraded Giant Metrewave Radio Telescope data REduction) package \citep{Kale2021}. Since we needed to obtain both Stokes I and V products for our science purposes, we modified the pipeline scripts.
In particular, since \texttt{CAPTURE} only uses the Stokes I model for self-calibration, which is not suitable for Stokes I and V imaging, we made separate RR and LL models and used them during the self-calibration steps.
We could perform self-calibration since there were several compact and bright enough sources on the field of view.
We first carried out four rounds of phase-only self-calibration with solution intervals of 8, 4, 2, and 1 min, followed by another four rounds of amplitude and phase calibration, using solution intervals of 4 2, and 1 min for the last two rounds, to obtain the final image at each observing epoch. 

We computed the dynamic spectra using the {\tt dftdynspec} function from the {\tt pwkit}\footnote{Available at \href{https://github.com/pkgw/pwkit}{https://github.com/pkgw/pwkit}} package \citep{Williams2017}. This function yields the dynamic spectrum of a point source by applying a discrete Fourier transform to its visibilities. Subsequently, we rebinned the dynamic spectra into frequency bins of 0.4~MHz to explore the presence of faint structures that might remain undetected with the native channel width (97.7~kHz). 

\section{Results} \label{sec:results}

\begin{figure}[htb!]
  \centering
  \includegraphics[width=9cm]{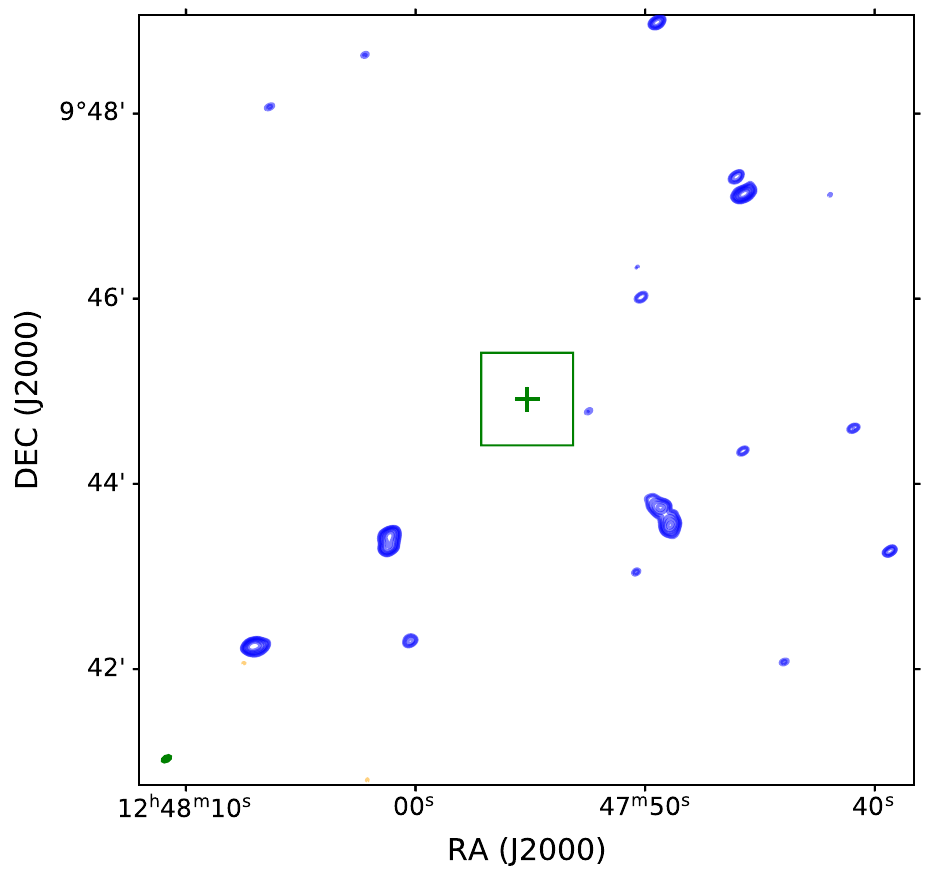}
\caption{\small Stokes I wide-field image centered on GJ 486 from band 4 (central frequency of 648 MHz) uGMRT observations on 30 October 2021.  The image covers a region of $\sim 8'\times8'$, centered at the position of GJ 486 (green cross).  The synthesised beam is shown as a green solid ellipse on the bottom left. The green square corresponds to a region of 60 $\times$60 arcsec squared, which we have used to estimate the local rms.}
\label{fig:30oct_images}
\end{figure}

\begin{figure}[htb!]
\begin{subfigure}{0.5\textwidth}
  \centering
 \includegraphics[width=\textwidth]{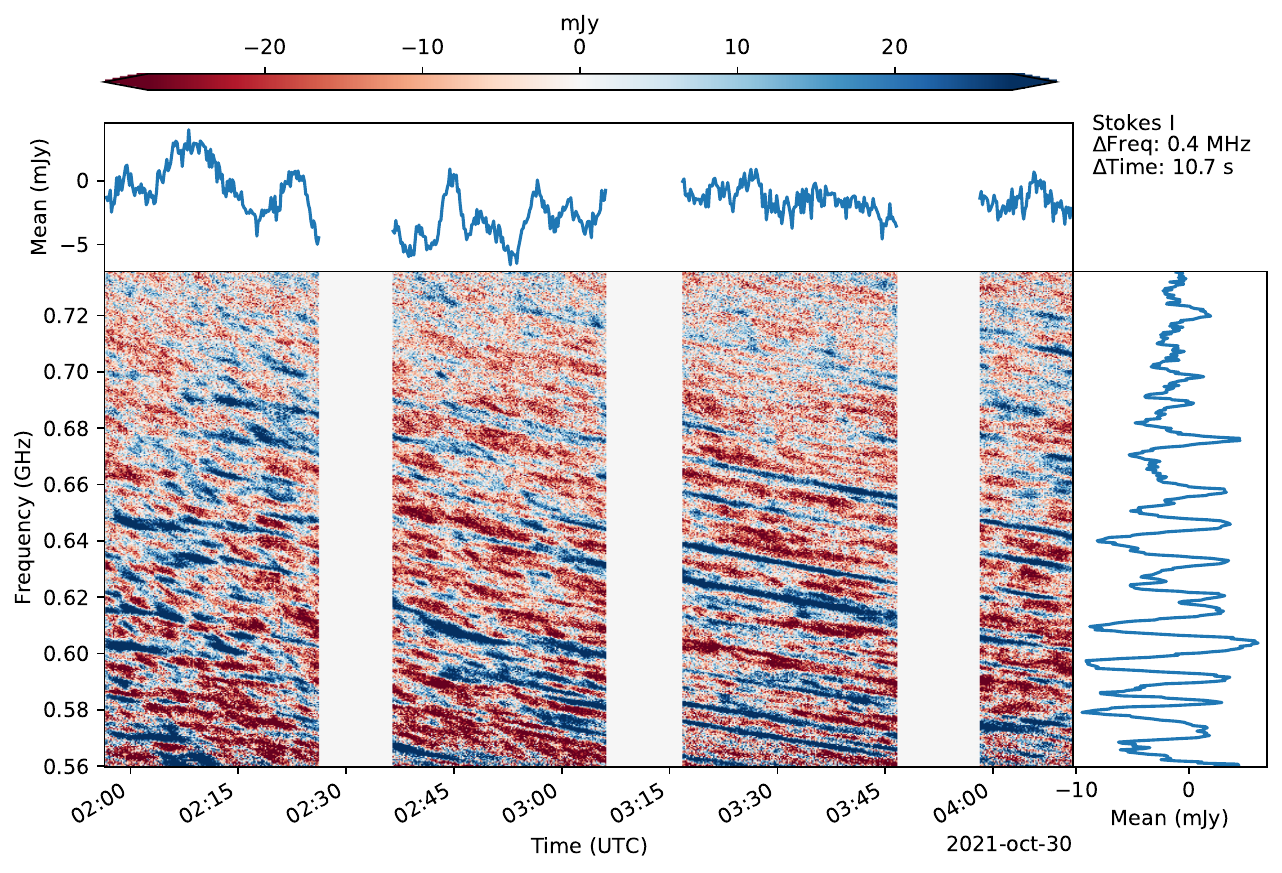}
\end{subfigure}

\begin{subfigure}{0.5\textwidth}
  \centering
  \includegraphics[width=\textwidth]{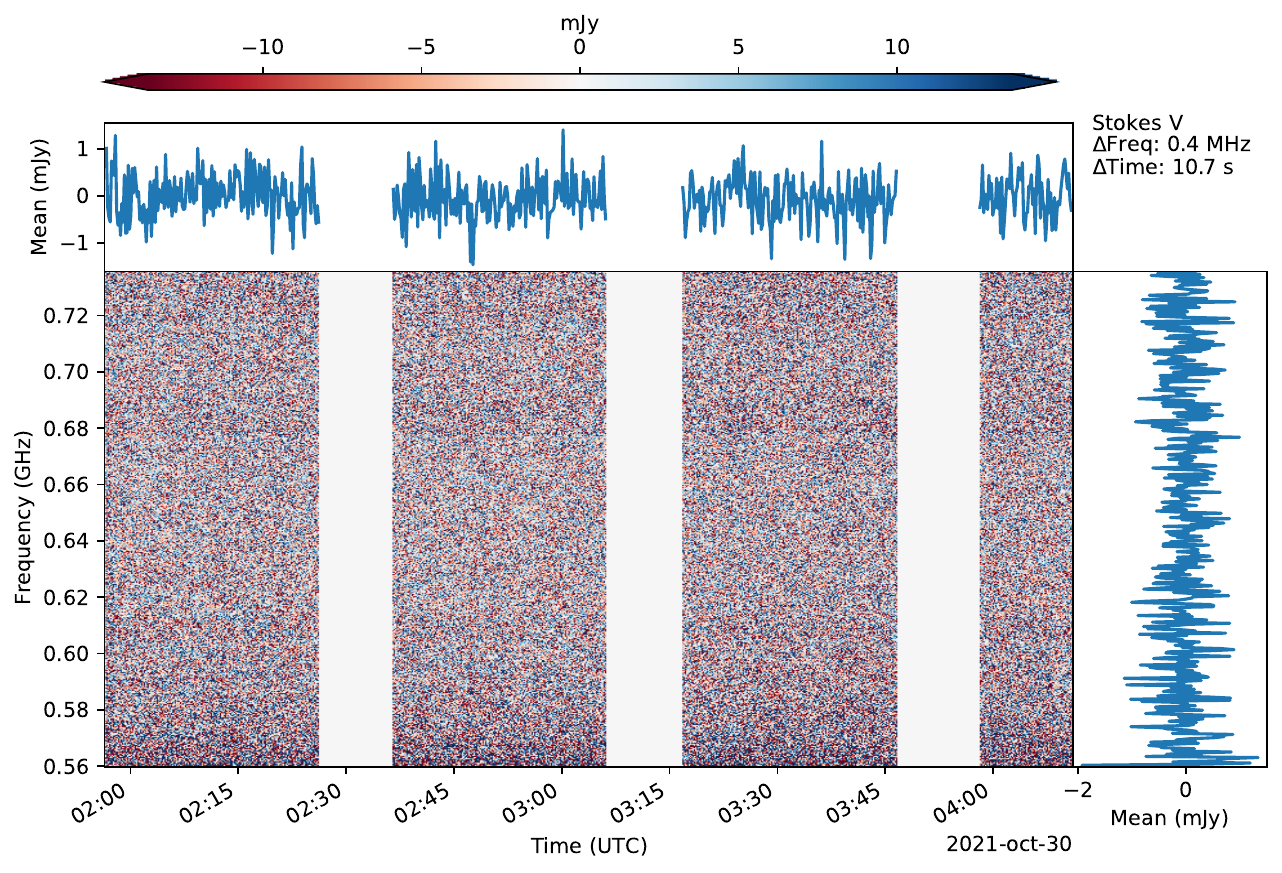}
\end{subfigure}

\caption{\small Dynamic spectrum of the Stokes I and V (upper and lower panels, respectively) emission from the GJ~486 -- GJ~486b system for our 30 October 2021 uGMRT observations in band 4, averaged in frequency ($\Delta\,\nu$ = 0.4 MHz). The time interval used is 10.7 s (which is the integration time). There is no apparent detection of radio emission above the noise. Blank regions correspond to times when we observed the phase calibrator. The spectra at the top and right of the dynamic spectrum result from collapsing all channels (top) and times (right). For a box located on time range 2:00 to 2:20 and frequency range 0.68 to 0.72 GHz, the rms values are 13.5 and 8.4 mJy for the Stokes I and V spectra, respectively.
}
\label{fig:dynspec}
\end{figure}

\noindent We summarize our main results in Table~\ref{tab:data} and
Figs.~\ref{fig:30oct_images} and \ref{fig:dynspec} (with additional information in Figure \ref{fig:ds_appendix}).  The main outcome is the non-detection of any steady
radio emission above 3$\sigma$ level, where $\sigma$ is the rms noise of the image, 
in any of our observing
epochs, neither in Stokes I, nor in Stokes V, at the position of \gj\
(Table~\ref{tab:data} and Fig.~\ref{fig:30oct_images}).  We calculated the
image rms within a 60 arcsecond-wide squared region centered on the position
of GJ~486 (green square in Fig. \ref{fig:30oct_images}). The rms value in the
images varied from epoch to epoch,  ranging from $17.6$ to $39.4\,\mu$Jy/b  and
from $9.9$ to $15.2\,\mu$Jy/b in Stokes I and V, respectively. Note that the
rms in the Stokes V images is significantly smaller than in the Stokes I
images, as the field is essentially depleted of sources emitting in Stokes V,
unlike in Stokes I. 
We also generated an image combining all the observing epochs, which also yielded non-detections. The rms of the stacked image is $16.7 \mu$Jy/b and $8.7$ $\mu$Jy/b for Stokes I and V, respectively.

\begin{table*}[ht!]
\caption{Observing log of the uGMRT observations of GJ~486
showing the value of local RMS in a region of 60 $\times$ 60 arcsecond centered on the source ($\sigma$).
}
\label{tab:data}
\begin{minipage}{\textwidth}
\begin{center}
\begin{tabular}{lllrrrrr}
\toprule
Observation date & Start time &  End time &  $\sigma_{\rm I}$  &  $\sigma_{\rm
V}$ &   bmaj & bmin & PA  \\
YYYY-MM-DD & HH:MM:SS.S &  HH:MM:SS.S &  ($\mu$Jy/b) &  ($\mu$Jy/b) &   (arcsec)& (arcsec)& (deg) \\
\toprule
2021-10-30 & 01:27:45.5 & 04:29:02.5 &     23.2 &  12.5 &  6.7& 4.1& -60.9 \\  
2021-11-07 & 04:02:54.9 & 07:52:31.0 &     22.2 &  13.0 &  4.8& 3.6& 30.4 \\  
2021-11-09 & 03:22:12.1 & 07:33:38.2 &     18.5 &  11.6 &  4.8& 3.5& 43.8 \\ 
2021-11-14 & 23:54:20.5 & 03:30:20.6 $^{+1}$&13.0 & 9.9 &  5.4& 3.9& 69.7 \\  
2021-11-17 & 00:00:34.4 & 03:32:59.7 &     22.6 &  11.5 &  5.9& 3.7& 61.1 \\  
2021-11-19 & 03:43:01.6 & 07:31:11.8 &     22.9 &  14.6 &  4.3& 3.6& 20.71 \\  
2021-11-20 & 04:30:51.6 & 08:27:37.2 &     22.5 &  10.5 &  5.3& 3.9& 56.9 \\  
2022-01-14 & 19:50:15.4 & 23:56:08.6 &     31.5 &  10.9 &  5.2& 3.8& -86.1 \\ 
2022-02-23 & 00:52:28.6 & 03:48:34.2 &     17.7 &  12.1 &  7.9& 3.8& 68.7 \\
\bottomrule
\end{tabular}
\tablefoot{The ``${+1}$''  indicates that the observations starting on 2021-11-14 ended the next day.}
\end{center}
\end{minipage}
\end{table*}

The rms values in Table \ref{tab:data}, and the images in Fig.
\ref{fig:30oct_images}, correspond to the averages over the entire observing
epoch, both in time and frequency. Hence, it could be possible that our observations might have
missed relatively bright, very short radio emission flares whose signal could
have been washed out by averaging the data. We therefore obtained the dynamic
spectra for all our observing epochs, both in Stokes I and V. The dynamic spectra confirmed that there was no bursting radio emission in any of our observing epochs. As an illustration, we show in Fig.~\ref{fig:dynspec} the  dynamic spectra
for our observations on 30 October 2021.  

While the dynamic spectrum of the Stokes I data seems to show some structure
on the emission, the dynamic spectrum of the Stokes V shows no signal across
the entire frequency and time domain, indicating  that, even if the features seen in the Stokes I dynamic spectrum were real, their signals were not circularly polarized, and therefore were not of an electron cyclotron maser origin. 
Most likely, the apparent structure in the Stokes I data is due to artifacts and/or sidelobe contributions from other sources that were not perfectly removed from the field after the 
\textit{uv}-subtraction.
We show in Fig. \ref{fig:ds_appendix} the rest of the Stokes V dynamic spectra.
As mentioned above, there is no sign of bursting emission in Stokes V, except for a single possible burst in the 20 November, which is due to an instrumental effect (see Appendix \ref{app:V-dynamic-spectra} for details).

\section{Discussion} \label{sec:discussion}

We discuss here the reasons that could have led to the non-detection of radio emission from magnetic star-planet interaction between GJ~486b and its host star, GJ~486, and its implications for the physical and geometrical parameters of the system. 
First, ECM emission is intrinsically time variable, so it is possible that nature did conspire against detecting it during our observations. However, our essentially full coverage of all orbital phases of GJ~486 makes this very unlikely.  We also note that since our observing campaign took place over several months, we cannot rule out that we could have missed bursting emissions between two observing epochs related to, e.g., coronal mass ejections (CMEs) impacting the planet. 
Second, we cannot exclude completely the possibility that the ECM fundamental frequency fell down below our observing frequency. In fact, since the relevant magnetic field for  ECM emission is the local magnetic field,  if the emission took place at a significant height above the surface of the star, the field may have been significantly smaller than the one initially estimated, and our observing frequency may have been too high to detect the emission at the fundamental frequency.  
We notice that although emission at the second harmonic could have been potentially detected by our observations, it is generally less intense than emission at the fundamental frequency, and rarely observed \citep{Melrose1982}. 

In the remaining of this section, we discuss
the implications of the non-detections due to a too low intensity of the radio emission, and/or to 
anisotropic radio beaming toward directions different from the observer's one. We address the former case by using a simple model to estimate the radio emission
expected to arise from the (sub-Alfvénic) magnetic interaction between the star and its
planet, which sets constraints on the stellar wind density at the orbit of GJ~486b, and
on its planetary magnetic field \ref{sec:results-model}. We address the latter scenario (anisotropic radio beaming) using the \texttt{MASER} code \citep{Kavanagh2023} to 
constrain the geometry of the system. \texttt{MASER} 
predicts the visibility of radio emission induced on a star via
magnetic star-planet interaction, as a function of time (Sect. \ref{sec:maser}). 

\subsection{Modelling the planet-induced radio emission from sub-Alfvénic interaction in GJ~486 -- GJ~486b}
\label{sec:results-model}

If a planet is in the sub-Alfvénic regime ($M_A = v / v_A < 1$.),  energy can be
transported upstream back to the star along the stellar magnetic field line connecting
to the planet \citep{Saur2013}. 
The key aspect, from an observational viewpoint, is that if this
energy is dissipated via the ECM instability, the frequency of the
emission should lie in the range of $\sim$100 MHz to GHz, as the relevant magnetic field is
that of the $B_\star$, which for M-dwarfs manifest with strengths ranging from $\sim$10 G
upwards to 1 kG \citep{Morin2008, Morin2010, Lehmann2024}, which correspond to cyclotron
frequencies in the above range. 

The energy powering the observed ECM (radio) emission comes from the Poynting flux,
$S_{\rm Poynt}$, generated at the orbit of the planet. More specifically, $S_{\rm Poynt}
\propto \, R_{\rm eff}^2\,v_{\rm rel}\,B_\perp^2$, where $R_{\rm eff}$ is the effective
radius of the planet, given by Eq.~\ref{eq:Rmp} below \citep{Zarka2001,Zarka2007} (or equal to the planet radius if the planet is not magnetized);  $v_{\rm rel}$ is the
relative velocity between the stellar wind flow and the planet;  and $B_\perp$ is the
component of the stellar wind magnetic field, perpendicular to the plasma velocity  at
the location of the planet.  

The total radio power emitted from one hemisphere of the star is $P_R = \beta\,S_{\rm
Poynt}$, where $\beta$ corresponds to the efficiency factor in converging Poynting
flux to ECM radio emission.  The value of $\beta$ is not known, but is expected to be 
in the range from 0.0001 to 0.01  for the planets of the Solar System \citep{Zarka2018b,Zarka2024}.
The radio flux density coming from the star can then be expressed as 

 \begin{equation}
   \label{eq:flux_density}
   F_R = \frac{P_R}{\Omega\,D^{2} \,\Delta\nu}, 
 \end{equation}
\noindent
where $\Omega$ is the solid angle into which the ECM radio emission is beamed, $D$ is the distance to the star from Earth, and $\Delta\nu$ is the total bandwidth of the ECM emission. Observations of the Io-DAM emission indicate that the beaming solid angle of a single flux tube is of about 0.16 sr \citep{Kaiser2000,Queinnec2001}. However, various reasons can lead to a larger solid beam of about a few times that value, e.g., a thick emission cone wall, a planetary plasma wake, or a distorted stellar magnetic field.  On the other hand, there is no reason to consider a full auroral oval around the star, which in the case of Jupiter yields a beam solid angle of 1.6 sr (see, e.g., \citealt{Zarka2004}). We therefore used a value of $\Omega$ = 0.5 sr in our simulations, corresponding to a few times the size of a single magnetic flux tube. 
We make the standard assumption that ECM radio emission has $\Delta\nu = \nu_{\rm g}$, where $\nu_g \approx 2.8\, B_\star$ MHz is the cyclotron frequency, and $B_\star$ is the average surface magnetic field strength of the star. 

We follow the prescriptions in Appendix B of \citet{PerezTorres2021} to estimate the observed radio emission arising from the sub-Alfvénic star-planet interaction in the GJ~486--GJ~486b system, and consider a closed dipolar geometry for the stellar magnetic field. We compute the predicted radio emission using the Zarka/Saur/Turnpenney model\footnote{
We note that in a number of previous works (e.g.,\citealt{Vedantham2020}, \citealt{Mahadevan2021}, \citealt{PerezTorres2021}), those models had been dubbed Zarka and Saur/Turpenney, since they were thought to predict significantly different levels of radio emission. However, during the Lorentz workshop ``The life cycle of a radio star'', held in Leiden, this issue was discussed and it was found that those models predict essentially the same flux densities, within about a factor of two \citep{Callingham2024,Zarka2024}.}

We approximate the stellar wind of GJ~486 as an isothermal, fully ionized, hydrogen plasma \citep{Parker1958}, which is characterized by the sound speed, or equivalently, the coronal temperature, $T_{\rm c}$. 
We use $\mathrm{log}\  T_{\rm c} = 6.67^{+0.40}_{-0.09}$ for GJ~486 from the work by \citet{Sanz-Forcada2024},  who utilized XMM-\textit{Newton} X-ray data to determine coronal temperatures for 21 CARMENES stars. 

The stellar wind density ($\rho_{\rm sw}$) at the orbit of GJ~486 b is not known beforehand. We therefore parameterize it via the stellar mass-loss rate, $\dot{M}_\star$:

\begin{equation}
    \label{eq:rho-wind}
    \rho_{\rm sw} = \frac{\dot{M}_\star}{4\pi\,d^2\,v_{\rm sw}},
\end{equation}

\noindent
which follows from the mass conservation equation for a time-independent stellar wind with constant \Mdotstar. Here, $d$ is the semi-major axis of the orbit of the planet (see Table~\ref{tab:gj486-params}), and $v_{\rm sw}$ is the stellar wind speed at the planet position. The stellar wind number density is 
$n_{\rm sw} = \rho_{\rm sw}/\mu m_{\rm p}$ and, since we assume a fully ionized, purely hydrogen plasma, $\mu = 1/2$.

The main difference with respect to the modeling in \citet{PerezTorres2021} is that here we also take into account the effect of free-free absorption. While for the nominal value of \Tc\ its impact is negligible, for the lower limit it can significantly impact the amount of transmitted radio-signal, if the stellar wind mass-loss rate is very high (see Fig. \ref{fig:SPI_transmitted}). We show the modeling of this effect in appendix \ref{app:free-free}.

\begin{figure}[htbp!]
\centering
\includegraphics[width=9cm]{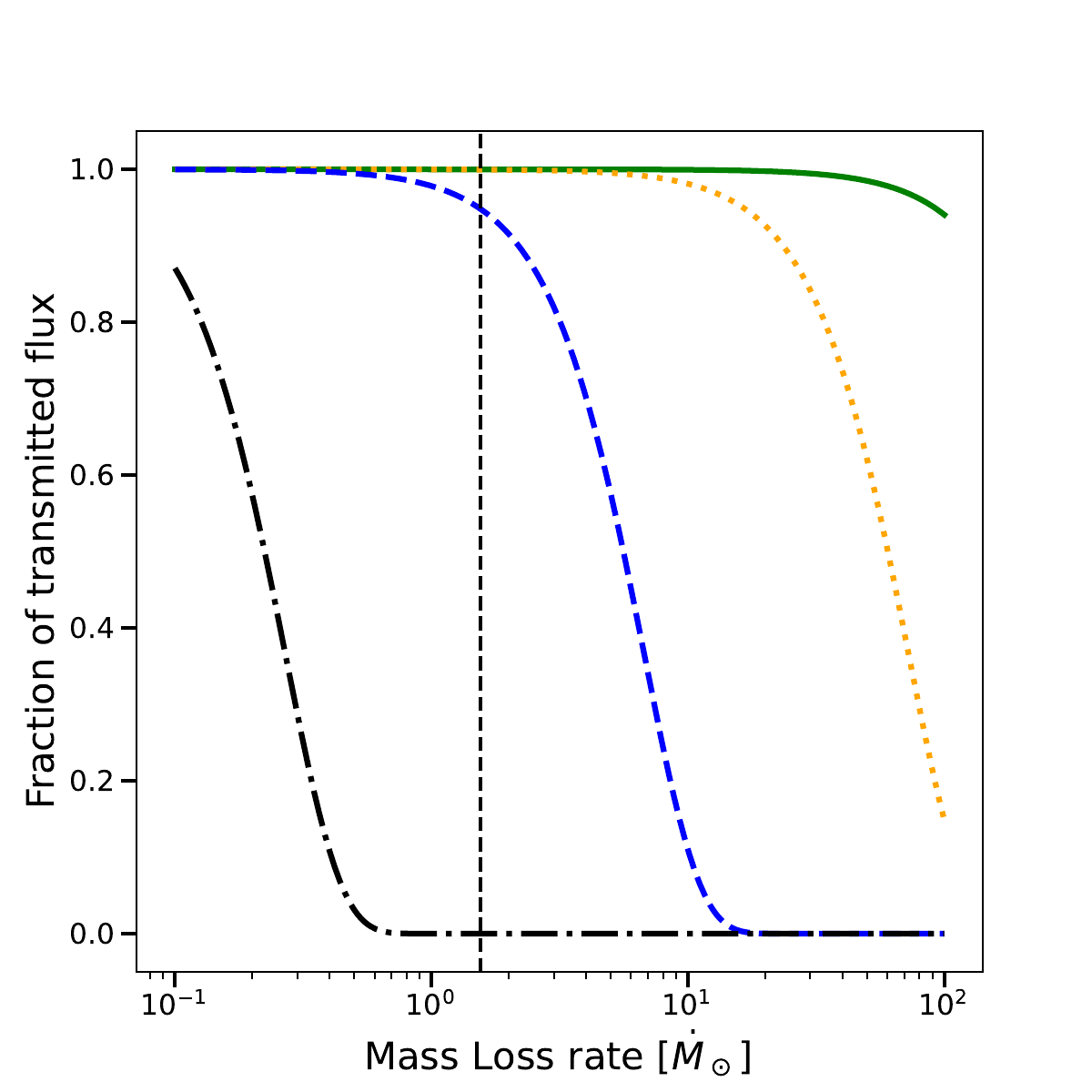}
\caption{\small 
Fraction of transmitted SPI flux due to the effect of free-free absorption for $T_c$ values of 1 (dot-dashed black line), 1.5 (blue dashed line), 2.5 (orange dotted line) and 4.7 (green solid line) MK. %(See text for details.)
}
\label{fig:SPI_transmitted}
\end{figure}

The effective obstacle radius of the planet is given by the magnetopause standoff distance of the planet, $R_{\rm mp}$, which we obtain by assuming a pressure balance between the stellar wind pressure (dynamic, thermal, and magnetic pressures) and the planetary pressure (magnetic pressure):

\begin{equation}
    \label{eq:Rmp}
    \Rmp = \kmp^{1/3} \Bigg[\frac{(B_{pl}/2)^2/8\pi}{ \pdynsw + \pthsw + \Bsw^2/8\pi} \Bigg]^{1/6} \Rp,    
\end{equation}
\noindent
where  $k_{\rm mp}$ is the factor by which the magnetopause currents enhance the magnetospheric magnetic field at the magnetopause, which is a value between 2 and 3. For simplicity, we adopt here $k_{\rm mp}$ = 2.  $B_{pl}$ is the intensity of the (dipolar) planetary magnetic field at the poles, \pdynsw\ = $\rho\, {\vrel}^2$ and \pthsw\ = \nsw $k_{\rm B} \Tc$ are the dynamic and thermal components of the stellar wind pressure, respectively; \Bsw\ is the magnetic field intensity of the stellar wind, and \Rp\ is the planetary radius. 
We note that Eq.~\ref{eq:Rmp} is an approximation to the actual value of the magnetopause standoff distance, and does not take into account effects such as the topology or the reconnection of the interplanetary magnetic field (IMF) with the magnetic field of the planet. For instance, the tilt of the magnetic field of the planet with respect to the IMF can increase the value of $\Rmp$ for a Proxima b-like planet by a factor of up to seven \citep{Penamonino2024}.

\begin{figure}[h!]
\centering
  \includegraphics[width=8cm]{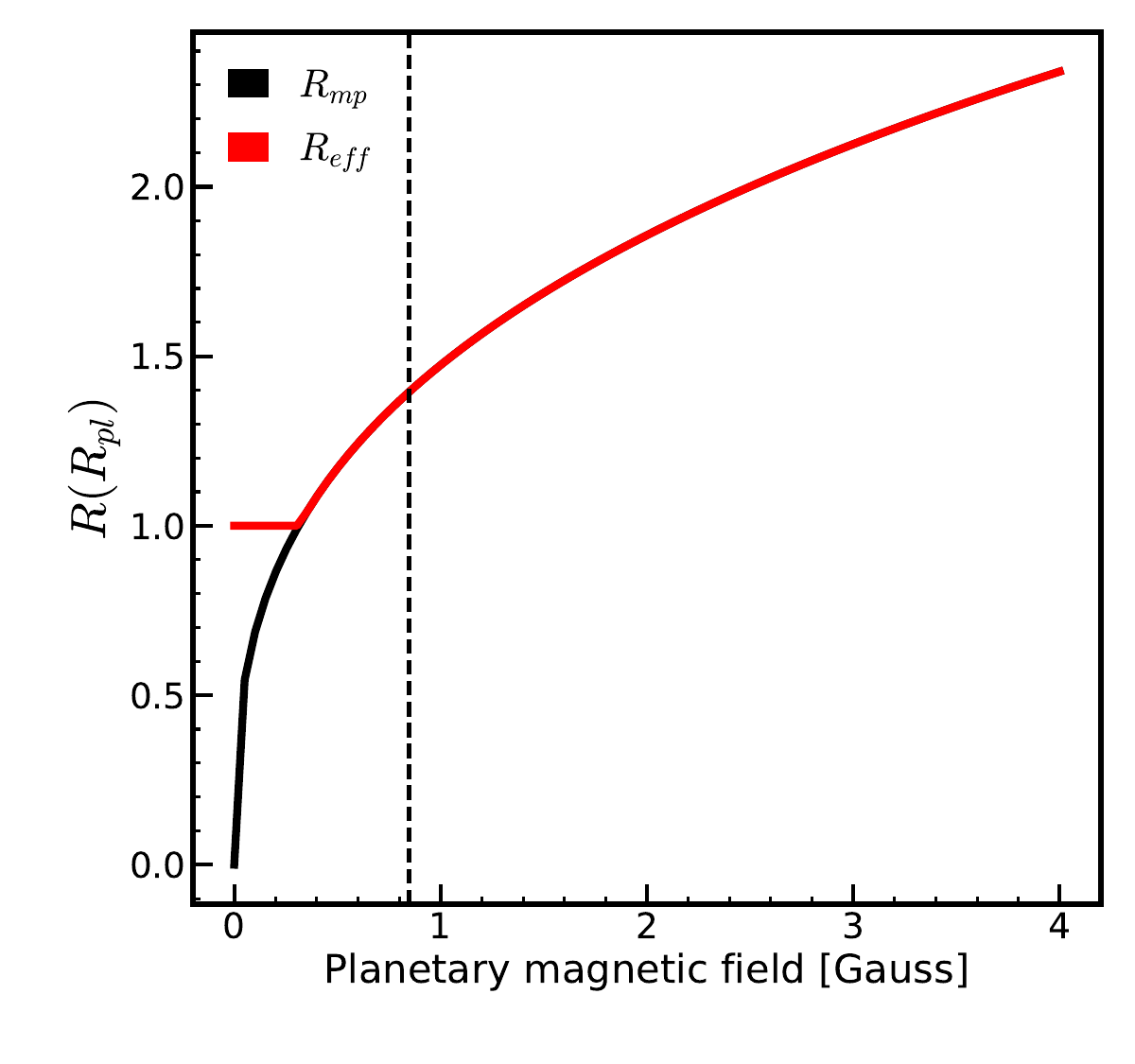}
\caption{\small 
 Magnetopause standoff distance, $\Rmp$, and effective radius, $R_{\rm eff}$, as a function of the magnetic field of the planet, \Bp. If \Bp\ $\gtrsim$ 0.35 G, $R_{\rm eff} = \Rmp$, otherwise $R_{\rm eff} = \Rp.$ The vertical dashed line corresponds to the nominal value of the magnetic field of 0.85 G.
}
\label{fig:R_eff_vs_Bp}
\end{figure}

Finally, we estimated the magnetic field of the planet using the Sano's scaling law  \citep{Sano1993}.
This resulted in a value of $B_{\rm pl}\simeq $0.85 G (see ~\ref{app:gj486b-field}),  around 40\% larger than the magnetic field of the Earth at the poles.
We show in Fig.~\ref{fig:R_eff_vs_Bp} the effective radius and magnetopause standoff distance as a function of the magnetic field of the planet.
If \Bp\ $\lesssim$ 0.35 G, then the effective radius is that of the planet itself. For \Bp = 0.85 G, the nominal value of the magnetic field, the effective radius is about 40\% larger that the physical radius of the planet.

%%%%
%%%% BEGINNING OF FIGURE FOR MODELLED RADIO EMISSION as function of M_dot_stellar_wind
%%%% 
\begin{figure*}[htbp!]
\centering

\begin{subfigure}{.33\textwidth}
  \centering
  \includegraphics[width=6cm]{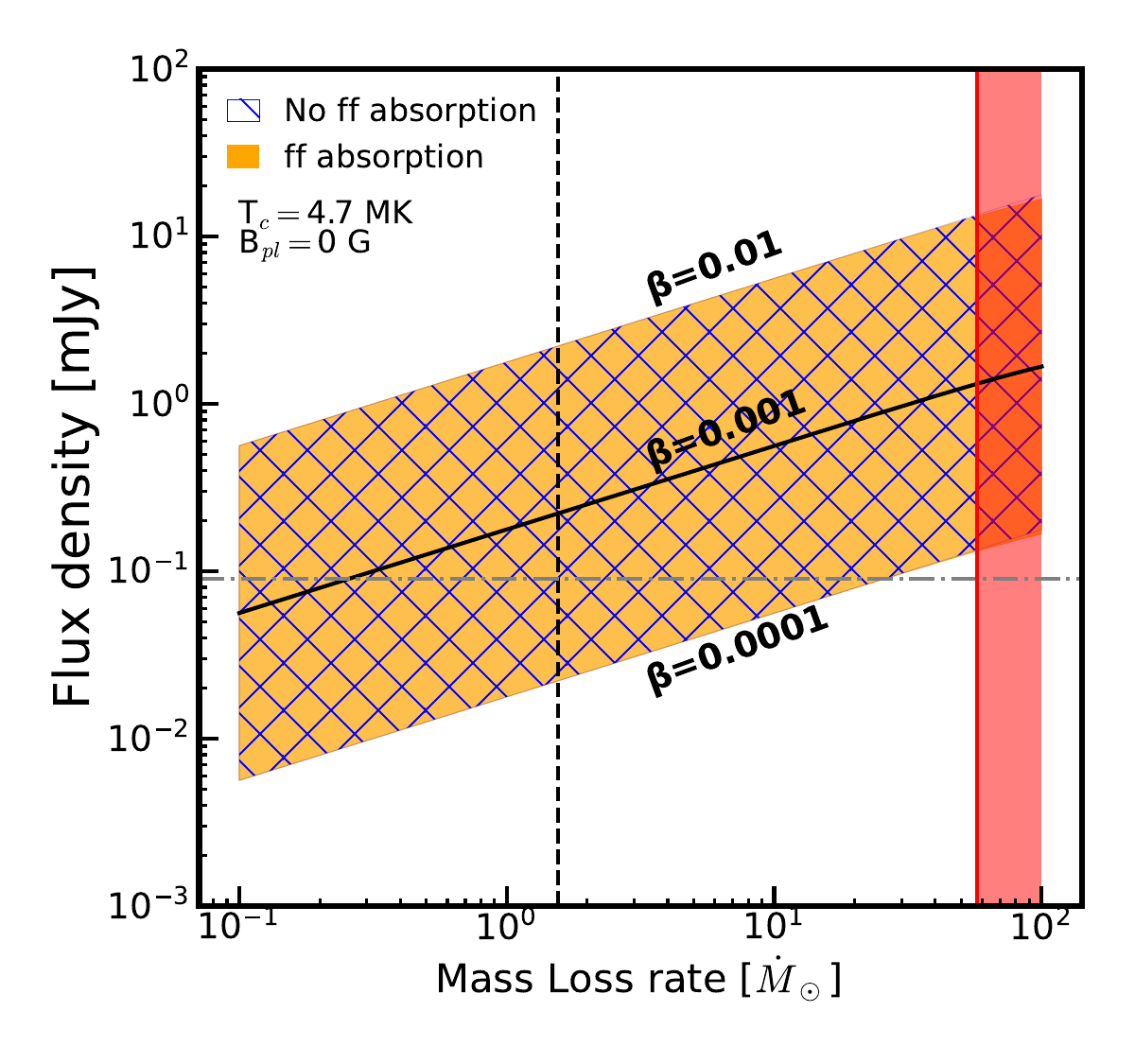}
\end{subfigure}%
\begin{subfigure}{.33\textwidth}
  \centering
   \includegraphics[width=6cm]{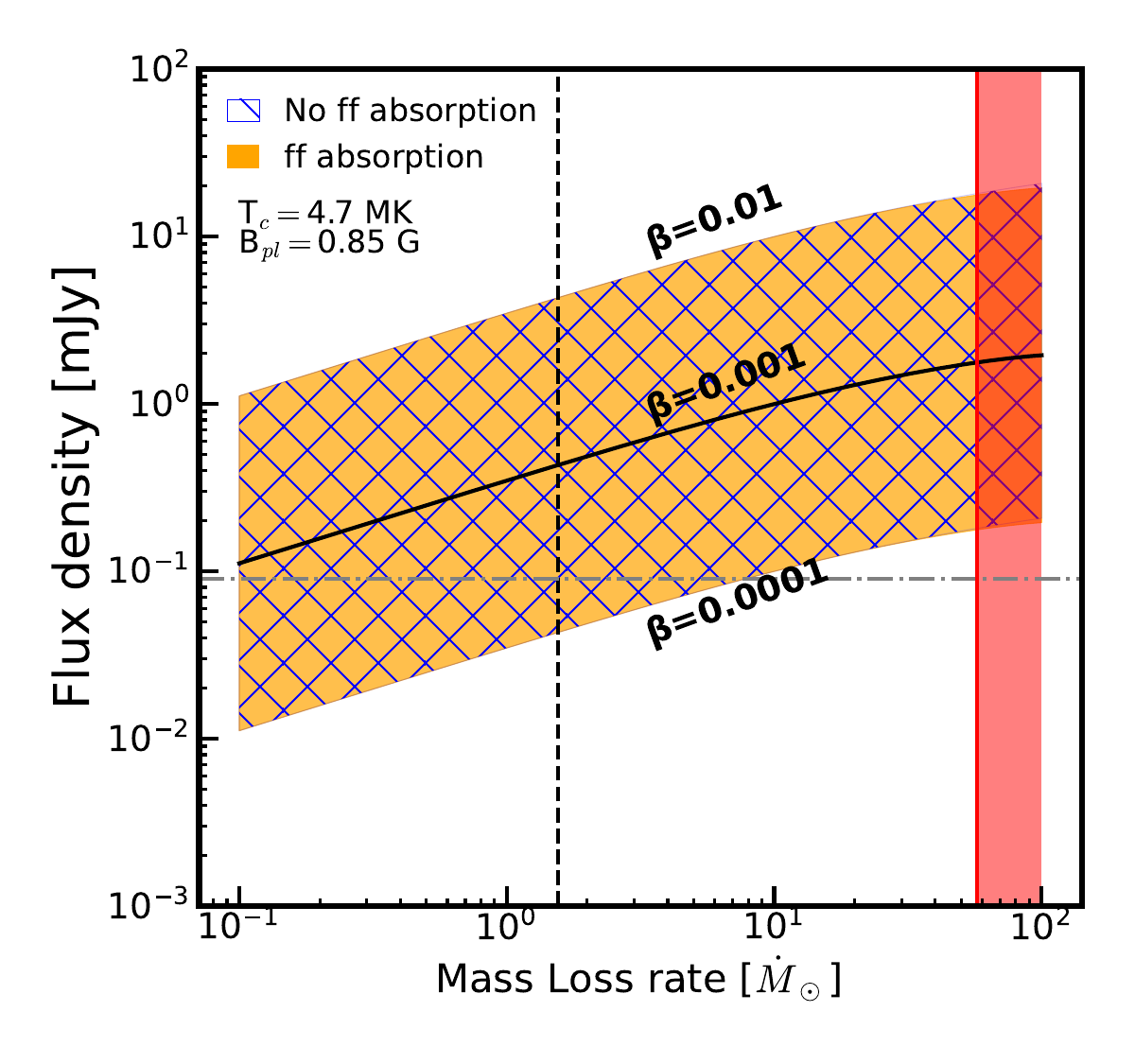}
\end{subfigure}%
\begin{subfigure}{.33\textwidth}
  \centering
  \includegraphics[width=6cm]{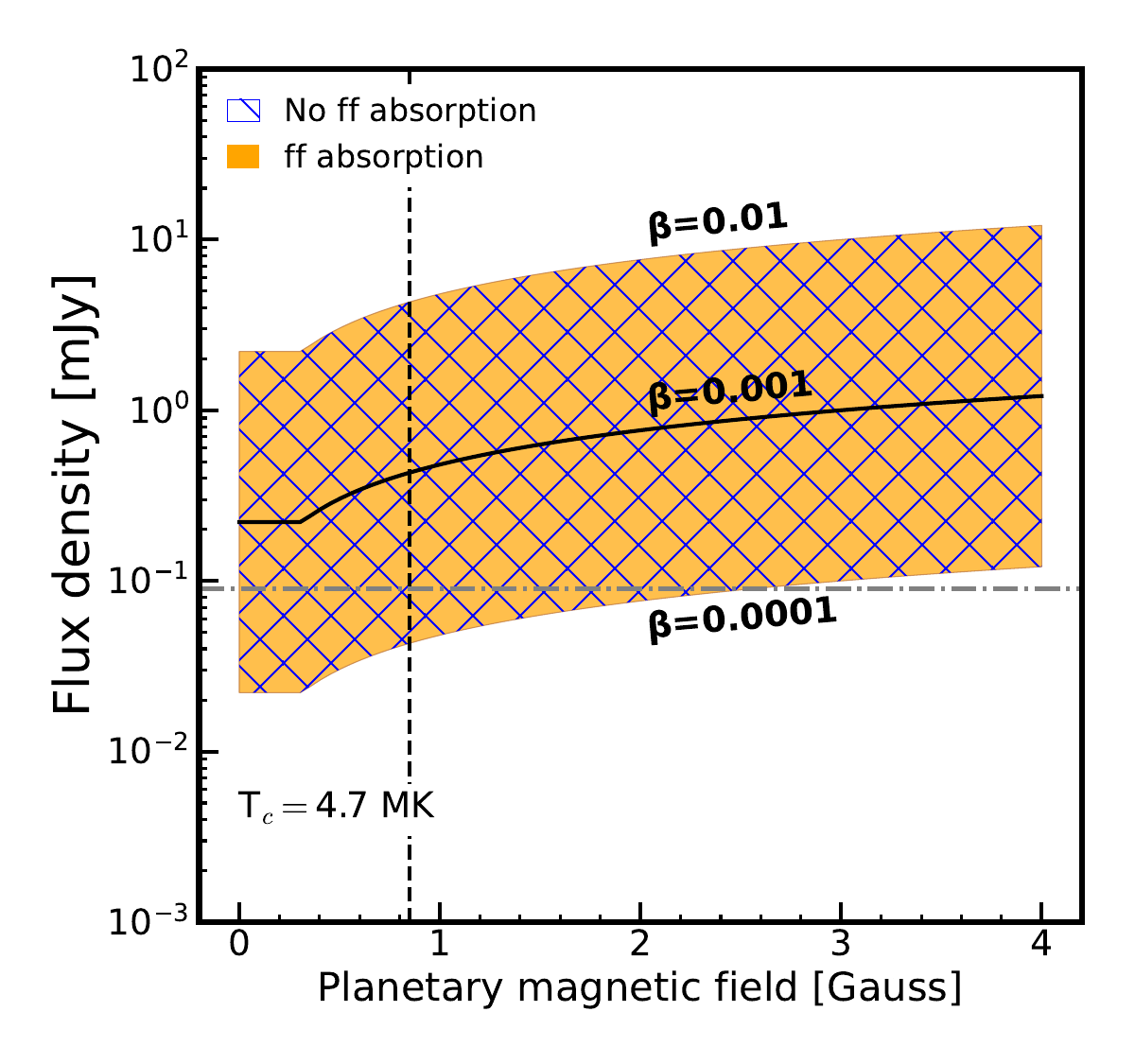}
\end{subfigure}%

\begin{subfigure}{.33\textwidth}
  \centering
  \includegraphics[width=6cm]{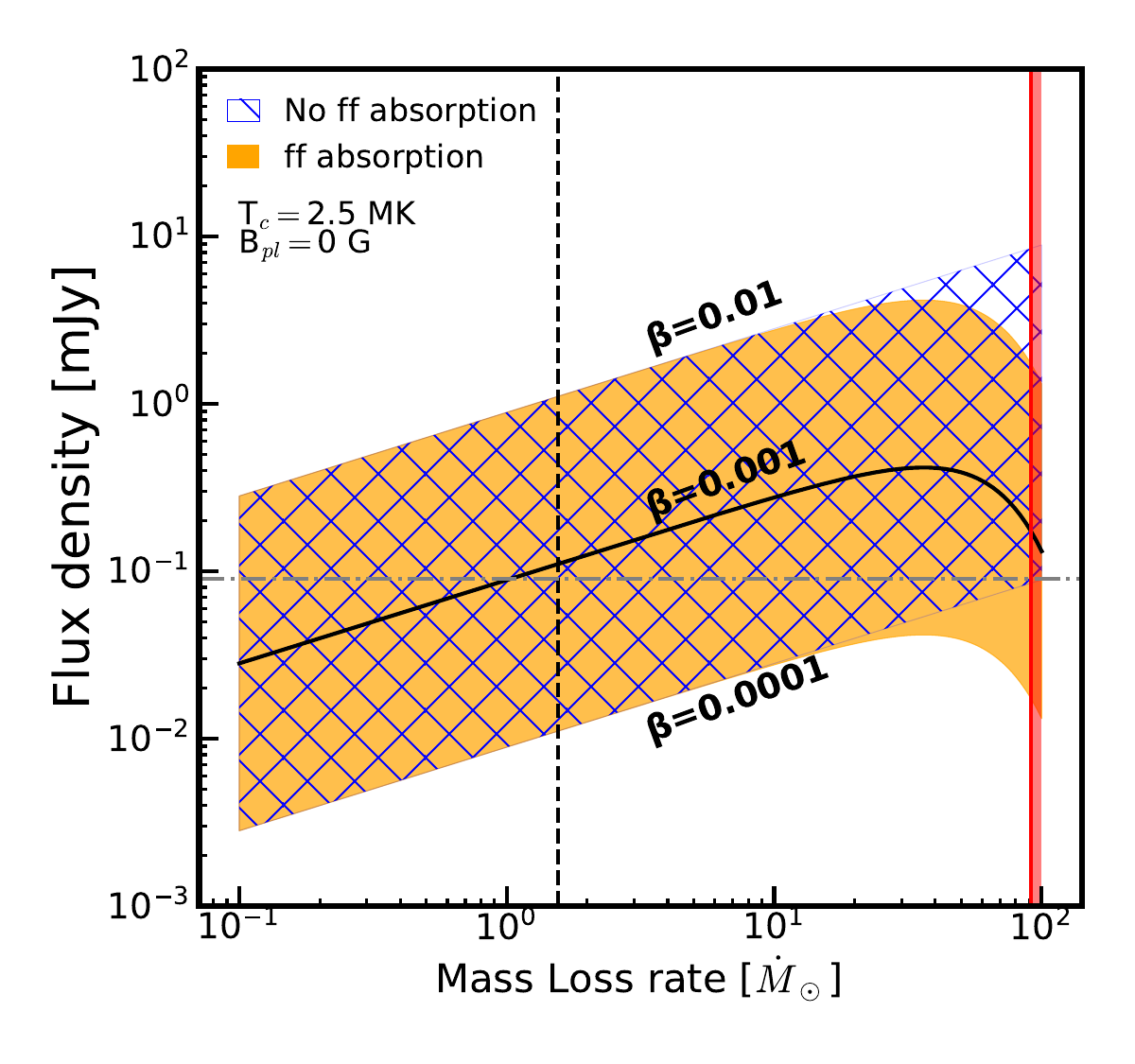}
\end{subfigure}%
\begin{subfigure}{.33\textwidth}
  \centering
   \includegraphics[width=6cm]{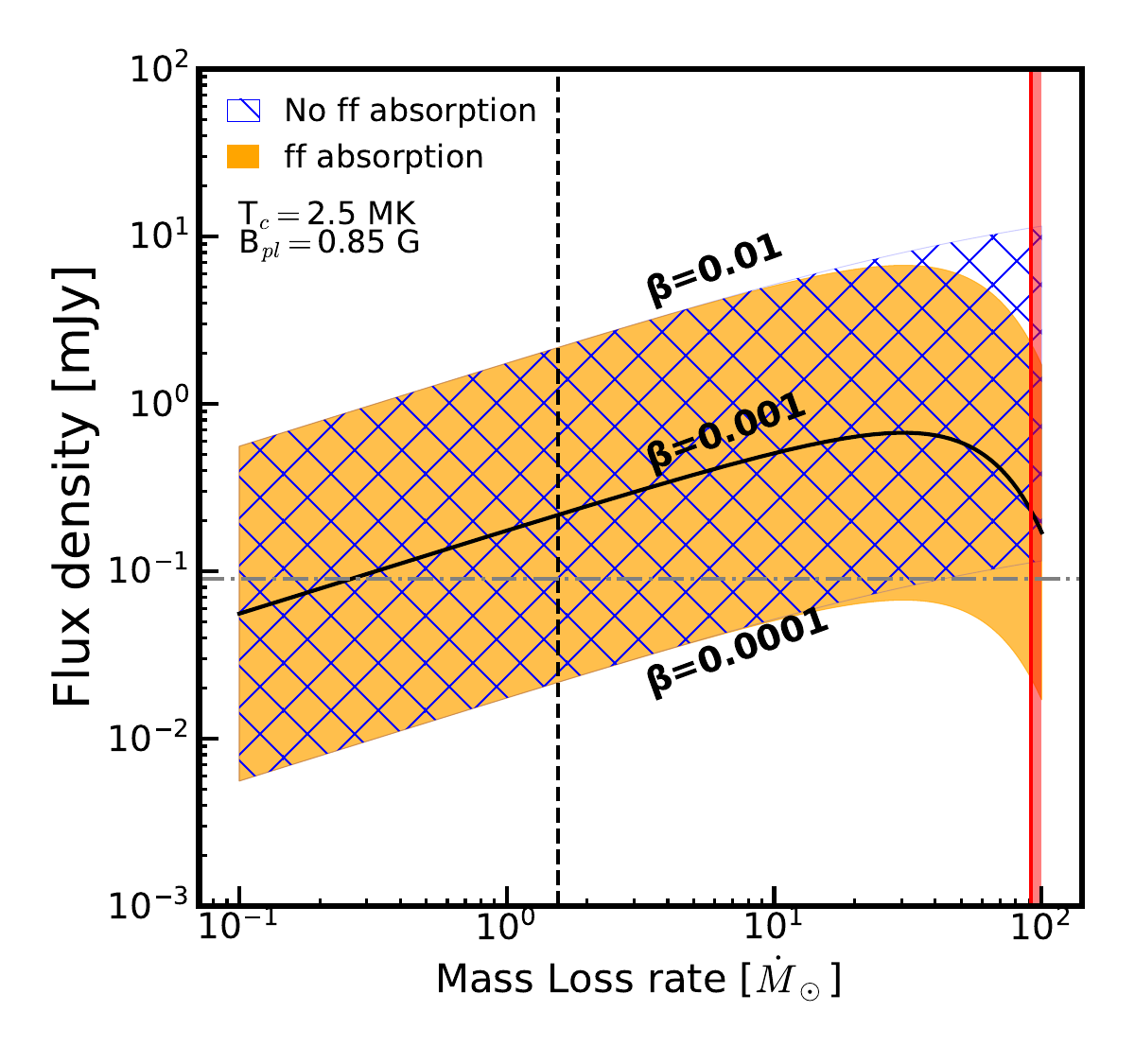}
\end{subfigure}%
\begin{subfigure}{.33\textwidth}
  \centering
  \includegraphics[width=6cm]{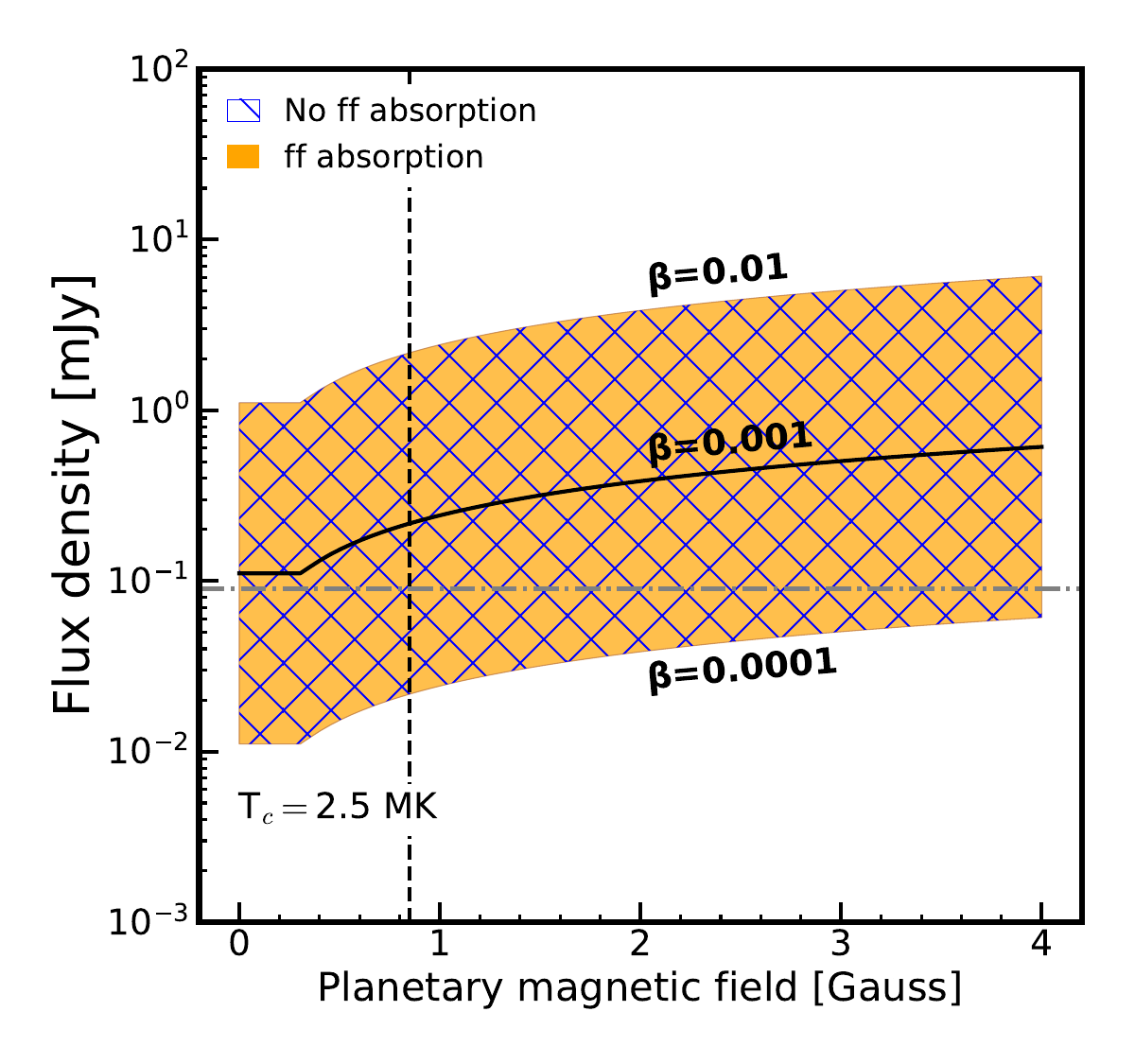}
\end{subfigure}%

\caption{\small
Predicted flux density arising at a nominal frequency of 670 MHz 
from star planet interaction. 
The yellow (cross-hatched) areas correspond to the expected radio flux density, including (neglecting) free-free absorption from the thermal electrons in the stellar wind.
The rows correspond to the expected emission for two different values of the coronal temperature, \Tc = 4.7 MK (the nominal value; top row)  and \Tc = 2.5 MK (3-$\sigma$ lower limit; bottom row).
The left and middle panels show the predicted flux density as a function of the stellar wind mass-loss rate for an unmagnetized planet (left) and a magnetized one with \Bp = 0.85 G (middle), with a vertical dashed line showing the nominal value, \Mdotstar\ = 1.4 \Mdotsun\ (see text for details). 
The right panels show the predicted flux density as a function of the planetary magnetic field, for the aforementioned nominal value of \Mdotstar, and a vertical dashed line corresponding to the nominal value of the magnetic field,  \Bp = 0.85 G. 
Red areas correspond to values $M_A \geq 1$, i.e., where the planet is in the Super-Alfvénic regime, therefore model predictions are not valid. 
The horizontal dash-dotted lines correspond to the 3$\sigma$ detection threshold of our observations  (where $\sigma = 30\,\mu$Jy b$^{-1}$).
}
\label{fig:SPI_flux}
\end{figure*}
%%%% END OF FIGURE FOR MODELLED RADIO EMISSION as function of M_dot_stellar_wind

\subsection{Constraints from the modeling of the radio emission}
\label{subsec:results-constraints}

In this section, we use the modeling described in Sect.~\ref{sec:results-model} to constrain some of the physical parameters of the system, in the case that the non-detections were due to the signal being too low to be detectable. 
We show in Fig.~\ref{fig:SPI_flux} the predicted radio emission arising from the sub-Alfvénic star-planet interaction in the GJ~486-GJ~486b system, as a function of \Mdotstar\ and \Bp\,  for two different values of the coronal temperature of GJ~486: $T_{\rm c} = 10^{6.67}=$ 4.7 MK (top), corresponding to the nominal value; and $T_{\rm c} = 10^{6.40}=$ 2.5 MK (bottom), corresponding to the 3-$\sigma$ lower limit of the coronal temperature determined by \citet{Sanz-Forcada2024}. 
Left panels correspond to the case of an unmagnetized planet, while middle ones correspond to the case of a planet with $B_{pl}$ = 0.85 G, the nominal estimated magnetic field value of GJ~486b (see Appendix \ref{app:gj486b-field}). 

The vertical lines in the plots correspond to the nominal value of \Mdotstar = 1.4 \Mdotsun , obtained by applying Eq. 7 in \citet{Johnstone2015b} to the GJ~486 star, and using the values in Table~\ref{tab:gj486-params}. 
Our predictions already take into account that the
efficiency of the conversion of Poynting flux into radio emission, $\beta$, is poorly known, but is very unlikely to be above 1\%, and possibly closer to 0.2\% or even lower \cite{Zarka2024}. 
This results in the yellow (cross-hatched) areas in the plots of Fig.~\ref{fig:SPI_flux}, which correspond to the expected radio flux density including (neglecting) free-free absorption from the thermal electrons in the stellar wind.

The majority of M dwarfs have winds that are weaker than, or comparable to, that of the sun, i.e., \Mdotstar $\lesssim$ \Mdotsun\ (\citealt{Wood2021}; \Mdotsun\ = 2$\times 10^{-14}$ \Msun\ yr$^{-1}$). However, \citet{Wood2021} found in their sample that two M dwarfs showed values of \Mdotstar = 30\,\Mdotsun\ (YZ CMi, M4 Ve) and  \Mdotstar = 10\,\Mdotsun\ (GJ 15AB; M2 V + M3.5 V). Therefore, in our study of the SPI radio-signal as a function of the stellar density, and given our lack of knowledge of the stellar mass loss rate of GJ~486, we used values of $\Mdotstar$ in the range from 0.1 up to 100 \Mdotsun.  This broad range encompasses the whole range of values inferred for M dwarfs to date \citep{Wood2021}. 

Our reference case is the one shown in the top middle panel of Fig.~\ref{fig:SPI_flux}, i.e., a magnetized planet with \Bp = 0.85 G orbiting its host star GJ~486, which has \Tc = 4.7 MK. 
The planet is in the Sub-Alfvénic regime as long as \Mdotstar $\lesssim$ 57 \Mdotsun, so the planet-induced Poynting flux can be transferred to the star and be re-emitted as auroral ECM emission at radio wavelengths. 
If the stellar wind mass-loss rate were higher, the planet would be always in a Super-Alfvénic regime, and there would be no emission arising from SPI. However, given the inferred value of  \Mdotstar\ for GJ~486, it is very unlikely that such mass-loss rates are at place.
Since the (nominal) value of \Tc\ is large, free-free absorption effects (yellow areas in the plots) are negligible, regardless of the value of \Mdotstar. 
For the nominal value of \Mdotstar = 1.4\Mdotsun, we should have clearly detected radio emission from SPI, independently of the value of $\beta$. Only if \Mdotstar\ is much smaller than \Mdotsun  (\Mdotstar $\lesssim$ 0.3 \Mdotsun) and if, at the same time, $\beta \ll 0.01$, could we have expected a non-detection. If the planet is not magnetized (top left panel), non-detections could be expected for values 
\Mdotstar $\lesssim$ 1.2 \Mdotsun, but also requiring small values of $\beta$.
The top right panel also shows that, for the nominal value of \Tc, we should have detected radio emission from SPI, whether the planet is magnetized or not, unless $\beta \lesssim 10^{-3}$.
The bottom panels show the same set of simulations as in the top panels, but for \Tc = 2.5 MK. The main differences with respect to the case of \Tc = 4.7 MK, are the following:  the planet enters the super-Alfvénic regime at much larger values of \Mdotstar ($\gtrsim 91$ \Mdotsun); the predicted flux is about a factor of two smaller; free-free absorption effects are noticeable, since \Tc\ is lower, albeit only for  large values of \Mdotstar. 
The strong lower limit on \Tc\ would have implied a detection of radio emission from SPI, if \Mdotstar 
$\gtrsim$ 1.2 \Mdotsun\ ($\gtrsim$ 4.7 \Mdotsun) for a magnetized (non-magnetized) planet, irrespective of the value of $\beta$. As in our reference case, non-detections imply not only very small values of $\Mdotstar$, but also that simultaneously the efficiency in converting Poynting flux into SPI radio emission is quite low ($\beta \ll 0.01$).
The bottom right panel shows essentially the same results as in the top right panel, i.e., that independently of whether the planet is magnetized, or not, we should have expected a detection of radio emission from SPI unless $\beta \lesssim 10^{-3}$.

Summarizing, if the non-detection of radio emission from SPI in GJ~486 was due to an intrinsically dim signal, this suggests that, independently of whether the planet is magnetized or not, the mass-loss rate must have been small (\Mdotstar $\lesssim$ 0.3 \Mdotsun) and that, concomitantly, the efficiency of the conversion of Poynting flux into radio emission was low, with values of $\beta$ close to 0.001, or even lower.
If the value of $\beta$ had been higher, then detections would have been warranted, regardless of the values of \Bp\ and \Mdotstar. 

\subsection{Constraints on the geometry of the system}
\label{sec:maser}

In the previous section, we discussed the constraints that can be inferred from modelling the radio emission, as described in Sect.~\ref{sec:results-model}, if the absence of a radio detection was due to the ECM signal being too faint to be detectable.   
Here we discuss an alternative scenario. Namely, that the signal might have been strong enough, but   
that our lack of knowledge about the stellar rotation, magnetic field geometry, and/or orbital and emission cone geometry could have resulted in the emission being beamed out of the line of sight during our observations.

We used the \texttt{MASER}  code developed by \citet{Kavanagh2023} to determine whether this was the case. \texttt{MASER}  takes in all key parameters pertaining to the geometry of the stellar rotation, stellar magnetic field, planetary orbit, and emission cone\footnote{The \texttt{MASER}  code assumes the emission cone properties are independent of the emission frequency. However, other existing codes such as ExPRES \citep{Hess2011, Louis2019} prescribe these properties based on the emission frequency, field strength, and velocity of the electrons powering the ECM emission. This however is informed by in-situ observations from Jupiter, which we lack in the case of GJ 486. Therefore, we opt to use \texttt{MASER} , which is also optimized for parallelized exploration of parameter spaces.}, and determines whether radio emission generated along the magnetic field line connecting the star to the planet is visible to the observer as a function of time (called a visibility lightcurve). We first constructed uninformed prior distributions for each parameter, which we list in Table~\ref{tab:maser priors}. From each distribution, we drew one million samples for each parameter, and computed the visibility lightcurve of the system during the observing windows listed in Table~\ref{tab:data}. (Note that we divided up each window into 100 points.) We then constructed a probability density for each parameter from the sets of samples that produced lightcurves with zero visibility. If any deviated from the assumed prior distribution, this implied that certain values for the parameter were more probable for the system. 

\begin{figure}
\centering
\includegraphics[width = \columnwidth]{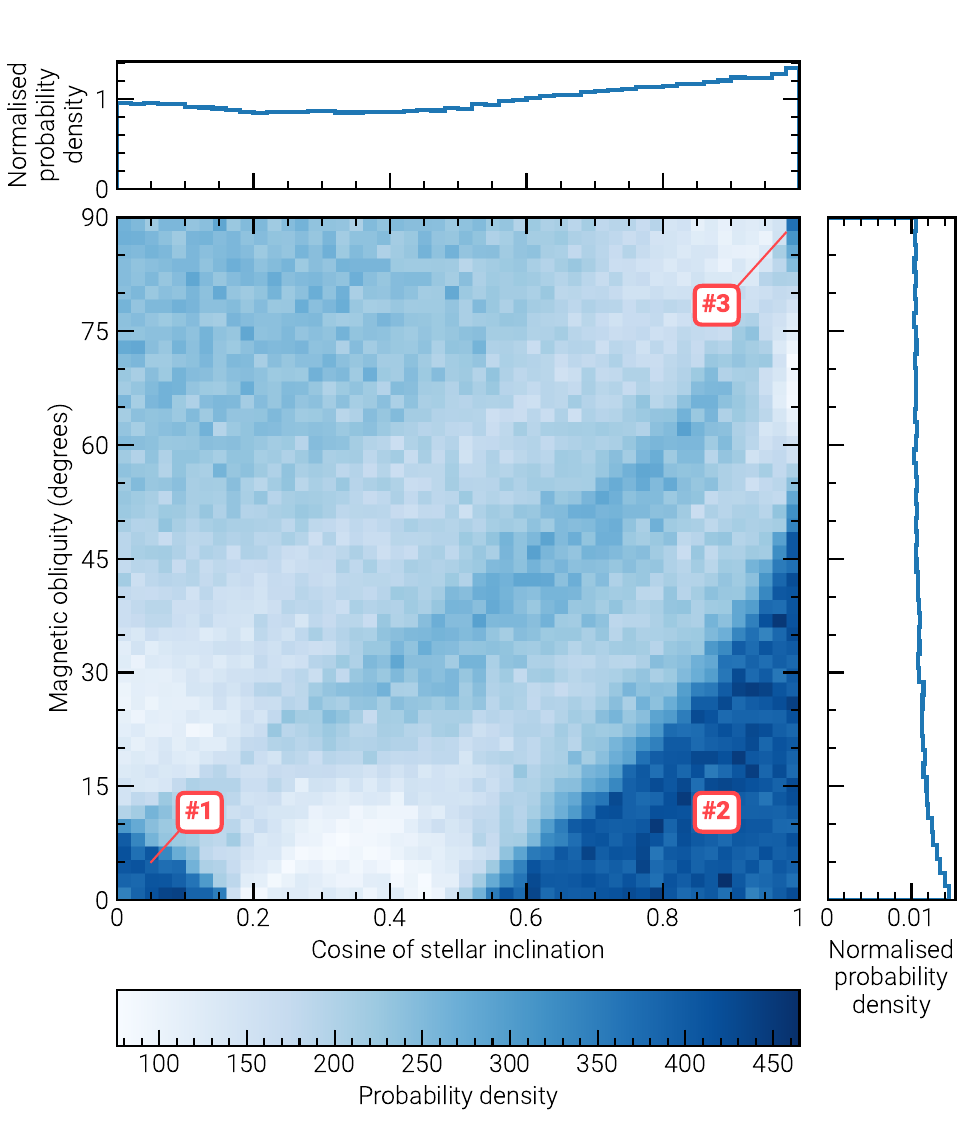}
\caption{\small Probability density for the stellar inclination and magnetic obliquity of GJ~486 given the non-detection of magnetic star-planet interactions. The darker blue colour corresponds to a higher probability of non-detection. The top and right panel shows the 1-dimensional histogram for each parameter. Note that this assumes that magnetic star-planet interactions take place in the system and we have the sensitivity to detect them (see main text for details).}
\label{fig:maser}
\end{figure}

\begin{figure}
\centering
\includegraphics[width = \columnwidth]{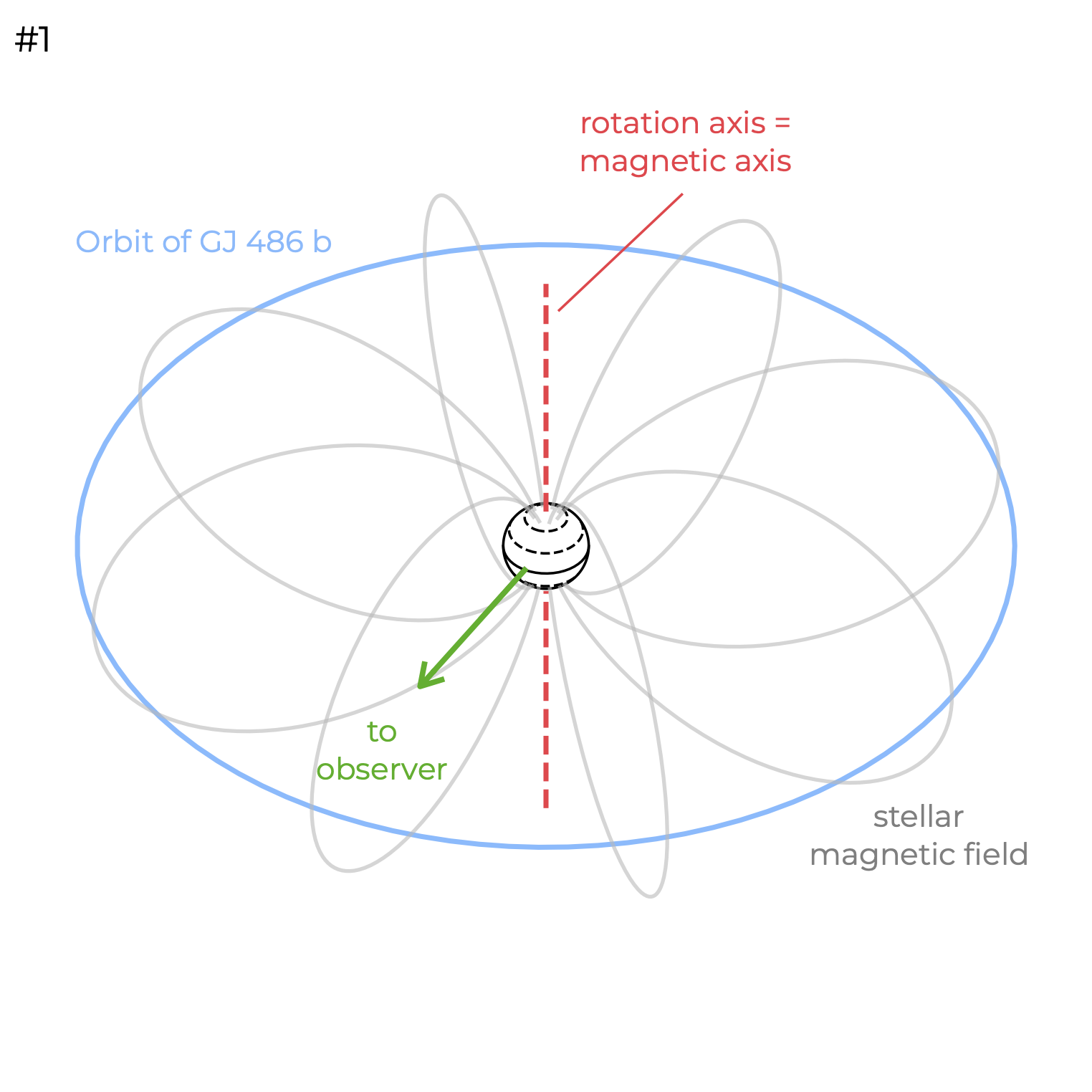}
\caption{\small To-scale sketch of the geometry for case \# 1, inferred for the GJ 486 planetary system based on our non-detection of star-planet interactions at radio wavelengths. The system is viewed equator-on, and has a low magnetic obliquity, meaning the magnetic dipole is aligned with the rotation axis (shown as a dot on the surface). The orbit of GJ 486 b is shown in blue, and the stellar magnetic field line connecting to the orbit is shown in red. The dashed lines on the stellar surface are lines of constant latitude, with the solid line showing the equator. Note that we find no bias towards any projected spin-orbit angle $\lambda$. For simplicity, we set $\lambda=0\degr$ here.}
\label{fig:geometry}
\end{figure}

\citet{Kavanagh2023} found that certain configurations for the stellar rotation and magnetic axis produce emission that is visible for a large percentage of time. These configurations describe systems where the magnetic axis of the star is always inclined relative to the observer by the cone opening angle, $\alpha$ (see Fig.~2 in \citealt{Kavanagh2023}). In the case of GJ~486 we find, however, the opposite. In Figure~\ref{fig:maser} we show a 2D probability density of the magnetic obliquity against the stellar inclination for systems with no visible emission. We see a pattern that is effectively the opposite of what is shown in Figure~8 of \citet{Kavanagh2023}, i.e. configurations where the magnetic axis never forms the angle $\alpha$ with the line of sight. 
Note that in Fig.~\ref{fig:maser}, the darker the color, the more likely is to have a non-detection of SPI in the  GJ~486 system (assuming SPI is in action in the system and we have the sensitivity to detect it).
We highlight the three most probable  configurations for a non-detection of SPI, denoted  \# 1, \# 2 and \# 3.
We show in Figure~\ref{fig:geometry} a sketch of the first possible configuration (\# 1). 
The region describes a configuration where the system is viewed edge-on, with a low magnetic obliquity. For completeness, we have also included to-scale sketches of the additional two most probable configurations inferred, \# 2 and \# 3, and show them in appendix \ref{app:geom_plots}. Those findings illustrate how non-detections of magnetic star-planet interactions can also be informative when combined with numerical models such as \texttt{MASER}.

\begin{table*}
\centering
\caption{List of unconstrained (or poorly constrained) relevant parameters of the GJ~486 system used in our \texttt{MASER}  simulations. }
\label{tab:maser priors}
\begin{tabular}{lcc}
Parameter & Prior distribution & Units \\ 
\toprule
Stellar rotation period ($P_\mathrm{rot}$) & $\mathcal{N}(49.9, 5.5)$ & days \\
Cosine of stellar inclination ($\cos i_\star$) & $\mathcal{U}(0, 1)$ & – \\
Stellar rotation phase at start of observations ($\phi_{\star,0}$) & $\mathcal{U}(0, 1)$ & – \\
Dipole field strength of the stellar magnetic field ($B_\star$) & $\mathcal{U}(100, 1000)$ & Gauss \\
Magnetic obliquity of the stellar magnetic field ($\beta$) & $\mathcal{U}(0, 90)$ & degrees \\
Projected spin-orbit angle ($\lambda$) & $\mathcal{U}(0, 360)$ & degrees \\
\bottomrule
\\
\end{tabular}
\tablefoot{\small 
For the prior distributions, we either use a normal ($\mathcal{N}$) or uniform ($\mathcal{U}$) distribution. For the normal distribution, the values in brackets are the center and standard deviation, and for the uniform distribution, the values are the minimum and maximum of the distribution. Note that we uniformly sample the cosine of the stellar inclination, which uniformly scatters the rotation axis on the surface of a sphere \citep[see][]{Kavanagh2023}. We fix the cone opening angle $\alpha$ to $75\degr$ and the cone thickness $\Delta\alpha$ to $1\degr$ in our simulation. Note also that extending the values for $i_\star$ and $\beta$ beyond $90\degr$ produces symmetric results of those shown in Figure~\ref{fig:maser}
}
\end{table*}

\section{Summary and outlook} \label{sec:summary}

We have presented the results of the longest radio monitoring campaign on the \gj\ - \gjb\ system, whose star hosts an Earth-like, rocky planet, aimed at finding evidence of radio emission arising from magnetic star-planet interaction. 
We observed \gj\ with the upgraded Giant Metrewave Radio Telescope (uGMRT) in the frequency range from 550 to 750 MHz (where we expected the cyclotron emission to take place, since this covers a surface magnetic field ranging from 196 to 268 G) in nine different epochs, between 30 October 2021 and 22 February 2022, covering almost all orbital phases
of \gjb.  
 
We obtained radio images of the region around \gj, both in Stokes I (total intensity) and V (circularly polarized intensity). We did not detect any quiescent radio emission in any epoch above a 3$\sigma$ noise floor of (33$-$96) $\mu$Jy/b and (30$-$45) $\mu$Jy/b  Stokes I and Stokes V images, respectively.
We obtained dynamic spectra in both Stokes I and V for all individual epochs, but did not find any evidence of short, bursty activity either. 
Our non-detections could have been due to four reasons: 
(1) time variability of the emission; (2) a different frequency range than the one we observed at; (3) an SPI radio signal too low to be detectable; and (4) the anisotropic radio beaming pointing to a direction away from the observer.

(1) Since we obtained an almost full coverage of the orbital phases of GJ~486b, and the resulting dynamic spectra showed no bursting signal at any time, it seems extremely unlikely that we missed the signal because of the putative SPI signal being time variable.  
(2) The frequency range where we observed was well tuned based on existing observations that indicate a magnetic field of about 240 G, although we cannot exclude that the maximum value of the cyclotron frequency fell below our lowest available frequency. This aspect suggests the need to either have beforehand stellar magnetic fields estimates from Zeeman Doppler Imaging (ZDI), or setting up simultaneous ZDI and radio observations. 

We tackled the potential issue of (3) the SPI radio signal being too faint by modeling the energetics of star-planet interaction (Sect.~\ref{sec:results-model} and \ref{subsec:results-constraints}). Our modeling strongly suggests that stellar wind mass-loss rate of GJ~486 must have been small (\Mdotstar $\lesssim$ 0.3 \Mdotsun), 
irrespective of whether the planet is magnetized, or not,  and that, simultaneously, the efficiency of the conversion of Poynting flux into radio emission, is very low ($\beta \lesssim 0.001$). If the value of $\beta$ had been higher, then detections would have been warranted, regardless of the values of \Bp\ and \Mdotstar. We also note that
free-free absorption effects are negligible, given the high value of the coronal temperature of the star (\Tc = 4.7 MK), and although they are noticeable  above $\Mdotstar \approx 20$ \Mdotsun for the strong lower limit of the coronal temperature of the star (\Tc = 2.5 MK), this should not have prevented the detection of radio emission from SPI.

Finally, we discussed the alternative possibility (4) that the SPI radio signal was strong enough to be detectable, but the  anisotropic radio beaming could have been pointing to a direction away from our line of sight, so we could have missed it. 
To this end, we used the \texttt{MASER}  code to constrain the geometry of the system. From our non-detections, we conclude that the magnetic obliquity and the stellar inclination are very low.

All of this illustrates how non-detections of magnetic star-planet interactions can also be informative when combined with a numerical modeling of both the energetic and the geometry of the system. 
We emphasize the need of accurate determinations of the  magnetic field strength of M stars hosting planets, using e.g., the ZDI method. ZDI  can also provide the stellar inclination and magnetic obliquity, and therefore would be an extremely useful tool to guide observing strategies at radio and other wavelengths.
We also emphasize the  need obtaining x-ray measurements of M dwarf stars to reliably determine their coronal temperatures, and assess the role of free-free absorption.

We end by noting that the detection of radio emission from star-planet interaction in M-dwarf systems hosting Earth-like exoplanets is proving to be very challenging, as the case of GJ~486 highlights, and the 
direct detection of small, Earth-like exoplanets would only be viable with hypothetical space-based instruments, like the FARSIDE initiative \citep{Hallinan2021}, among others.
On the other hand, the detection of Jupiter-like planets from radio observations is more promising, as their magnetic fields are expected to be large enough as to have their associated gyrofrequency above the ionosphere cut-off, and the radio emission should be much larger than that from an Earth-like planet. However, the only potential detection is so far that of $\tau$ Boo \citep{Turner2021}.  While efforts at the extreme edge of low-frequencies ($\nu \lesssim 50$ MHz), like NenuFAR, are worthwhile, current low-frequency ($\nu \lesssim 100$ MHz) radio interferometers lack the necessary sensitivity and are affected by severe radio frequency interference to detect directly radio emission from Jupiter-like exoplanets, we may still need to wait for LOFAR 2.0 and SKA-low to be operational.

%%%%%%%%%%%%%% ACKNOWLEDGMENTS %%%%%%%%%%%%%%
\begin{acknowledgements}
We thank the referee, Philippe Zarka, for his thorough and insightful review, which significantly improved our manuscript.
We thank Sanne Bloot for her suggestion to take into account free-free absorption effects in our modelling of the radio emission.
LPM, MPT, GB, JFG, JM, GA, AA, PA, DR, MO, and JAC acknowledge financial support through the Severo Ochoa grant CEX2021-001131-S,  and through 
the Spanish National grants PID2023-147883NB-C21,
PID2020-114461GB-I00, PID2023-146295NB-I00, and PID2022–137241NB–C42, 
all of them funded by MCIU/AEI/ 10.13039/501100011033,
LPM also acknowledges funding through grant PRE2020-095421, funded by MCIU/AEI/10.13039/501100011033 and by FSE Investing in your future.  We also acknowledge the service and support of the Spanish Prototype of an SRC (SPSRC), funded by the Spanish Ministry of Science, Innovation and Universities, by the Regional Government of Andalusia, by the European Regional Development Funds and by the European Union NextGenerationEU/PRTR.
G.B-C acknowledges support from grant PRE2018-086111, funded by MCIN/AEI/ 10.13039/501100011033 and by 'ESF Investing in your future' 
\end{acknowledgements}

%%%%%%%%%%%%%% REFERENCES %%%%%%%%%%%%%%
\bibliographystyle{aa} % style aa.bst
\bibliography{bibfile} % your references biblio.bib

%%%%%%%%%%%%%% APPENDIX %%%%%%%%%%%%%%%%%%%%%%
\begin{appendix}

\section{Planetary magnetic field} 
\label{app:gj486b-field}

Since the planet GJ~486b has a bulk density slightly larger, but close to being Earth-like, we estimated its magnetic field by assuming a Sano scaling law \citep{Sano1993}, which is adequate for magnetized planets of the Solar System.

The magnetic field moment, $\mathcal{M}$ is then  $\mathcal{M}\, \propto\, \rho_c^{1/2}  \, \omega \, r_c^{7/2}.$ 
Here, $\rho_c$ is the mass density in the dynamo region, $\omega$ is the rotational speed of the planet, and $r_c$ is the planet core radius, assumed to be
$\sim 0.55 \, R_{\rm pl}$, as in the case of the Earth. 
For simplicity, we assume $\rho_c$ to be the planet bulk density, which we estimate by using
the planetary mass and radius in Table~\ref{tab:gj486-params}.
Since the rotational speed, $\omega$, of GJ~486b is unknown, we assume that GJ~486b is tidally locked, so that its rotational speed is equal to the orbital speed.
The surface magnetic field of GJ~486b is then $\Bp\ = \mathcal{M}/ \Rp\ $. Adopting $B_\Earth = 0.5$ G, we get \Bp\ = 0.85 G.

\section{Free-free absorption}
\label{app:free-free}
Free-free absorption attenuates the radio signal by a factor of $e^{-\tau_{\nu}}$, where the optical depth, $\tau_{\nu}$, is defined as \citep{Cox2000}

%, which is proportional to $T_{\rm c}^{-1/2}\, \nu^{-3} \, n_e\, n_i$. 
\begin{equation}
    \label{eq:ff_tau}
    \tau_\nu =  \int_{R_\star}^{ \infty } \kappa_\nu \,dz ,
\end{equation}
\noindent
We integrated along the line of sight from the site where the emission takes place, which we conservatively took as the stellar surface, to the observer (at infinity). In practice, we integrated up 10$^4$ $R_\star$, at which point the effect of free-free absorption became negligible in all cases. The coefficient $\kappa_\nu$ is (e.g. \citealt{Cox2000})

\begin{equation}
    \label{eq:ff}
    \kappa_\nu = 3.692 \times 10^{8} \left(1 - e^{-h \nu / k_B T}  \right) Z^{2}\, g T^{-1/2}\, \nu ^{-3}\, n_e\,n_p \, \bigr[\mathrm{cm^{-1}}\bigr],
\end{equation}
\noindent
where $\nu$ is the frequency of the observed emission, $h$ is the Planck constant, $k_B$ is the Boltzmann constant, $Z$ is the ionization state (+1 for a fully ionized hydrogen medium), $T$ is the temperature of the medium (which in this case is $T_c$, since  we assume an isothermal wind), and 
$n_e$ and $n_i$ are the electron and ion number density, respectively. All magnitudes are in cgs units. 
Since we have a fully ionized hydrogen plasma, $n_i = n_p = n_e$, where $n_p$ is the proton density.

Finally, $g$ is the Gaunt factor, which in the radio regime is given by \citep{Cox2000}:

\begin{equation}
    \label{eq:gaunt}
    g = 10.6 + 1.9  \log _{10} (T) - 1.26 \log _{10} (Z \nu ) .
\end{equation}

%%%%%%%%%%%%%%%% 
\section{Other geometric configurations yielding non-detections of SPI}
\label{app:geom_plots}

\begin{figure}[h!]    
  \centering
\includegraphics[width=0.4\textwidth]{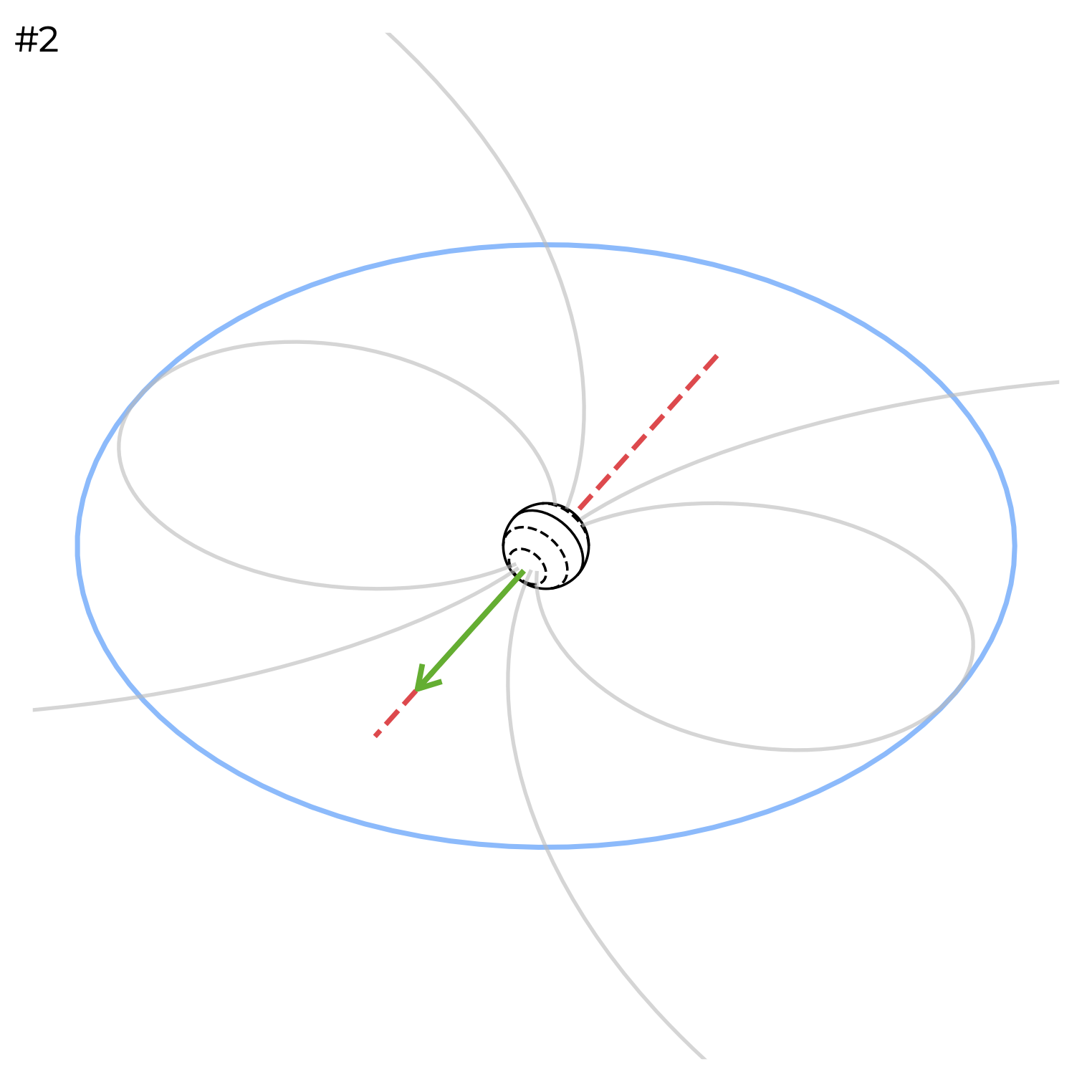}
  \caption{\label{fig:config2} \small Sketch of the geometry \# 2 inferred for the GJ 486 planetary system, based on our non-detection of star-planet interactions at radio wavelengths. The system is viewed pole-on, and has a low magnetic obliquity.} 
\end{figure}

\begin{figure}[h!]    
  \centering
\includegraphics[width=0.4\textwidth]{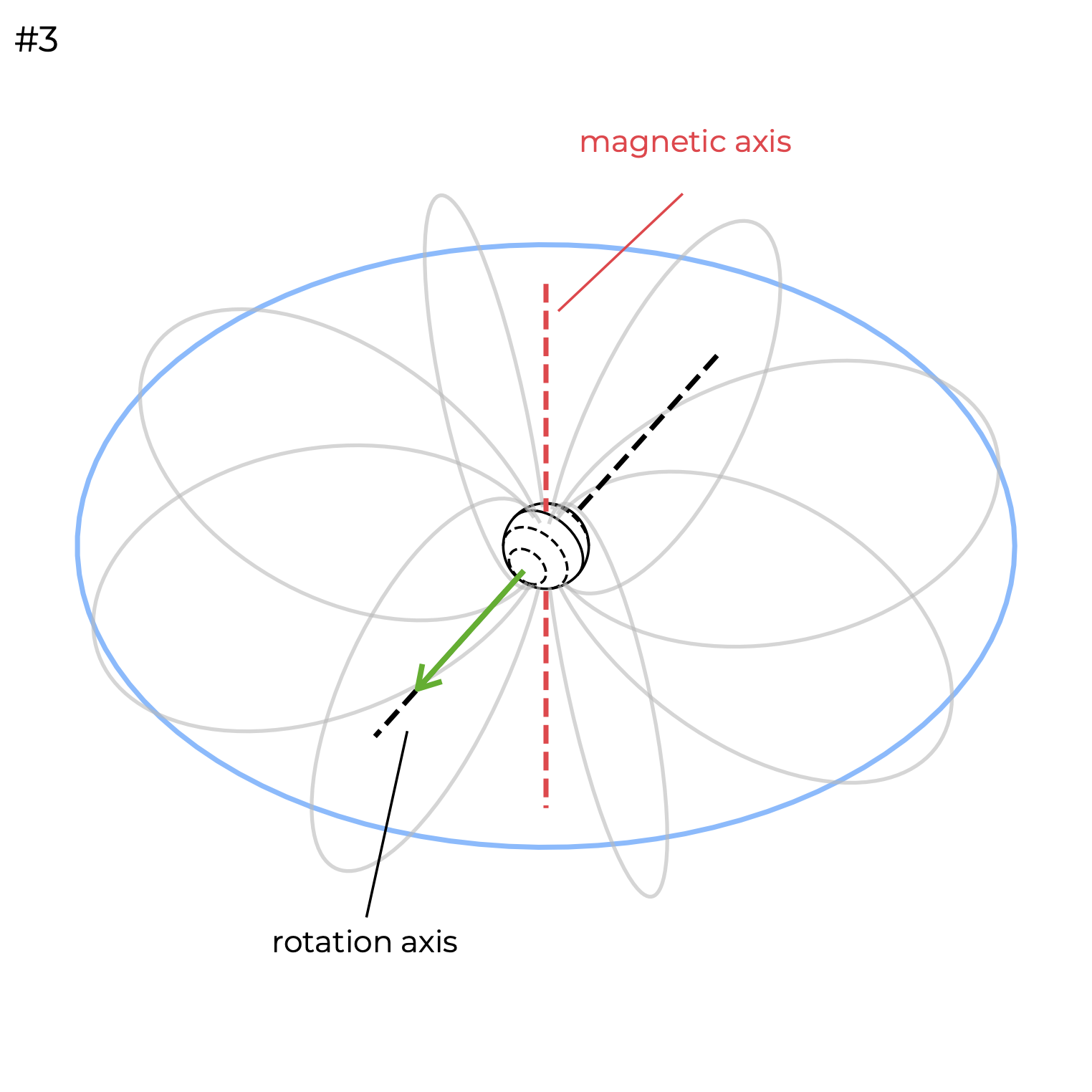}
  \caption{\label{fig:config3} \small Same as in Fig.~\ref{fig:config2}, but for our system viewed pole-on and with a magnetic obliquity  close to 90$^\degree$ (geometry \# 3).} 
\end{figure}

We show in Figs. \ref{fig:config2} and \ref{fig:config3} sketches for the alternative geometric configurations \# 2 and \# 3 depicted in Figure \ref{fig:maser}. Those configurations maximize the probability of non-detection of SPI (see Sect.~\ref{sec:maser} for details).

%%%%%%%
\section{Stokes V dynamic spectra} 
\label{app:V-dynamic-spectra}
We show here all Stokes V dynamic spectra for all observing epochs, but the 30 October 2021, which is shown in Fig.~\ref{fig:dynspec}. Note that all dynamic spectra are featureless, indicating that there is no bursting emission. The only possible exception is a burst-like signature around 07:00 hr on 20 November 2021. To determine whether this feature corresponded to a real emission, we 
obtained the dynamic spectra using 
four different phase centers. The resulting spectra had always the same feature seen in the original dynamic spectrum (centered at the position of \gj\ ), implying that this bursting feature is not real, but due to an instrumental effect.

\begin{figure*}[p!]    
\begin{subfigure}{0.45\textwidth}
  \centering
  \includegraphics[width=8cm]{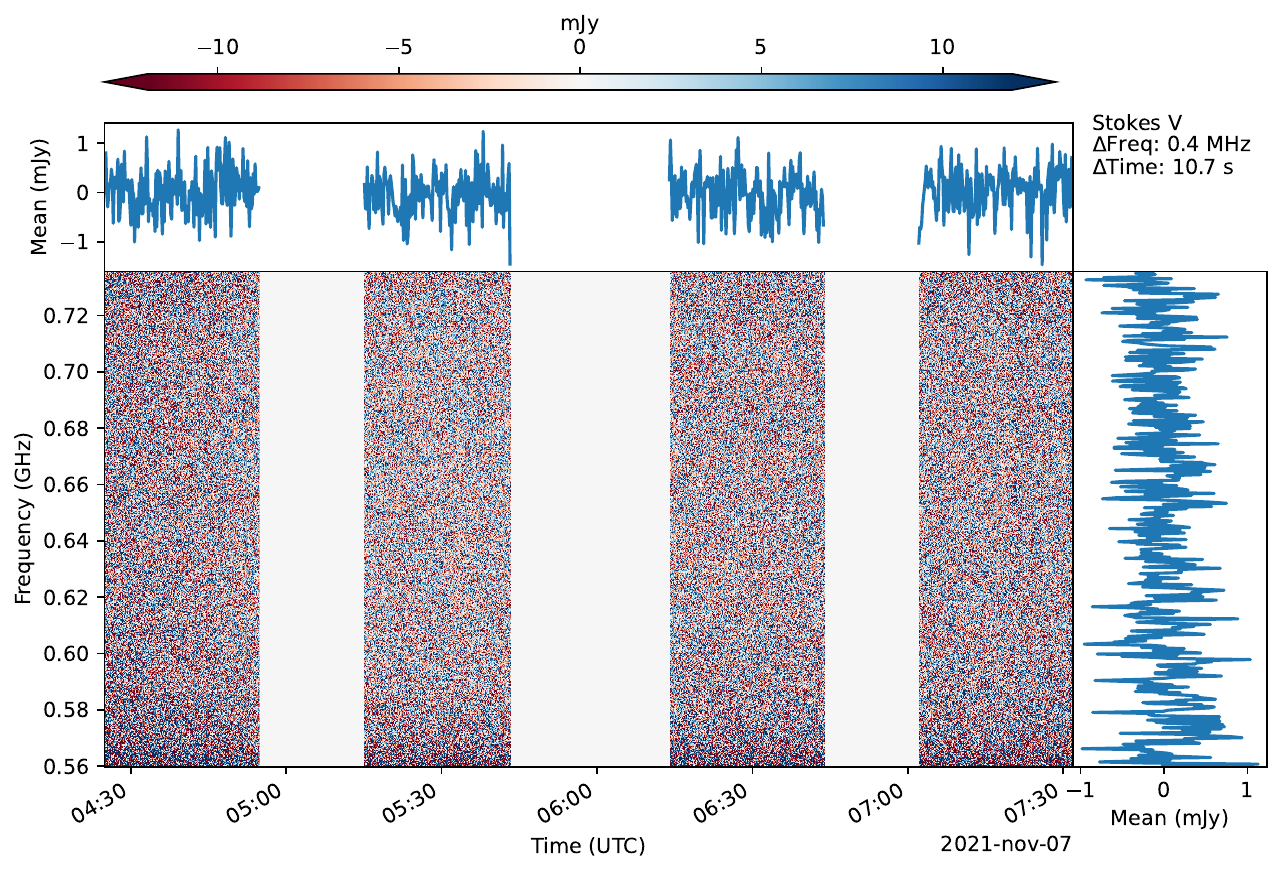}
  \caption{\label{fig:07nov2021_stokes_V_dynspec} \small Stokes V dynamic spectra for 7 November 2021. } 
\label{fig:07nov_dynspec}
\end{subfigure}
\begin{subfigure}{0.45\textwidth}
  \centering
  \includegraphics[width=8cm]{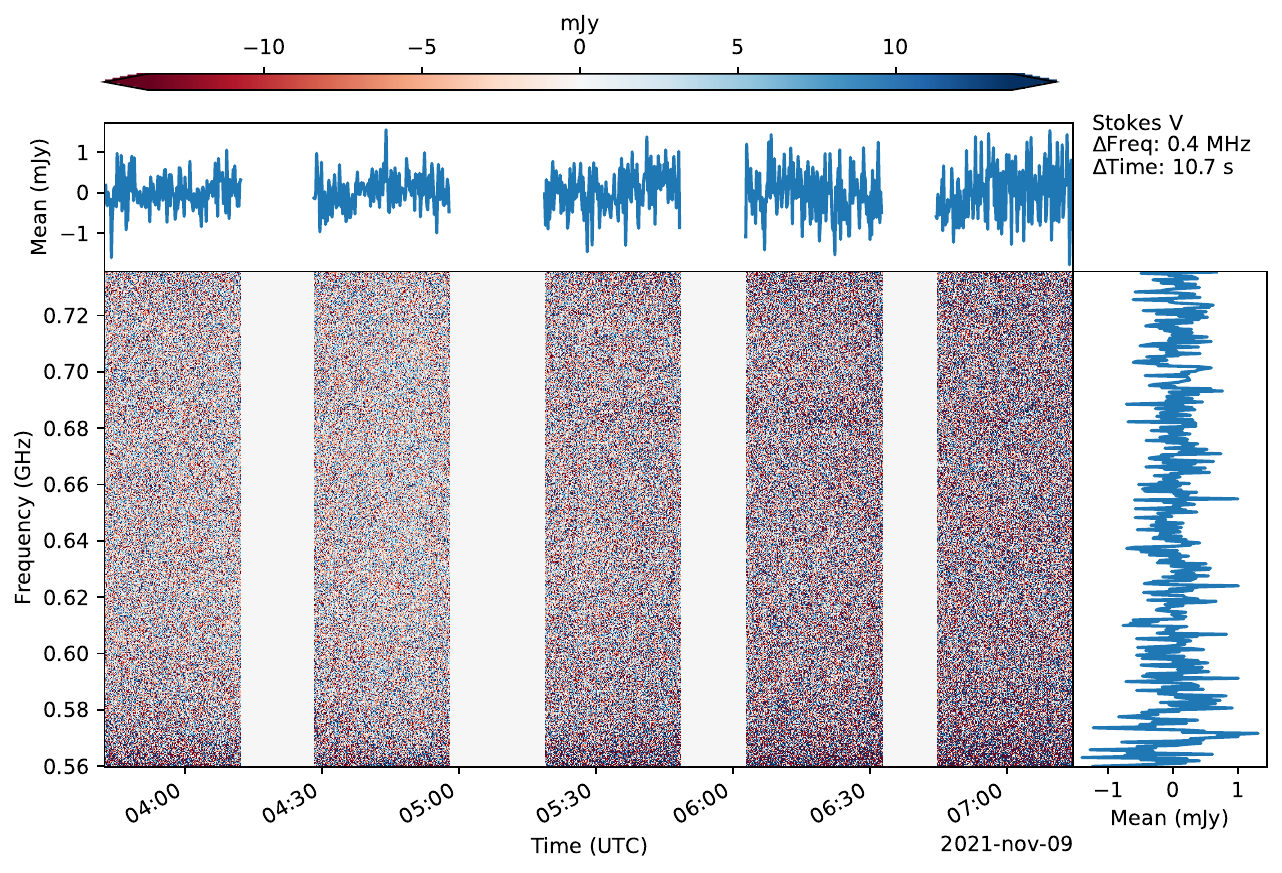}
  \caption{\label{fig:09nov2021_stokes_V_dynspec} \small Stokes V dynamic spectra for 9 November 2021. } 
\label{fig:09nov_dynspec}
\end{subfigure}

\begin{subfigure}{0.45\textwidth}
  \centering
  \includegraphics[width=8cm]{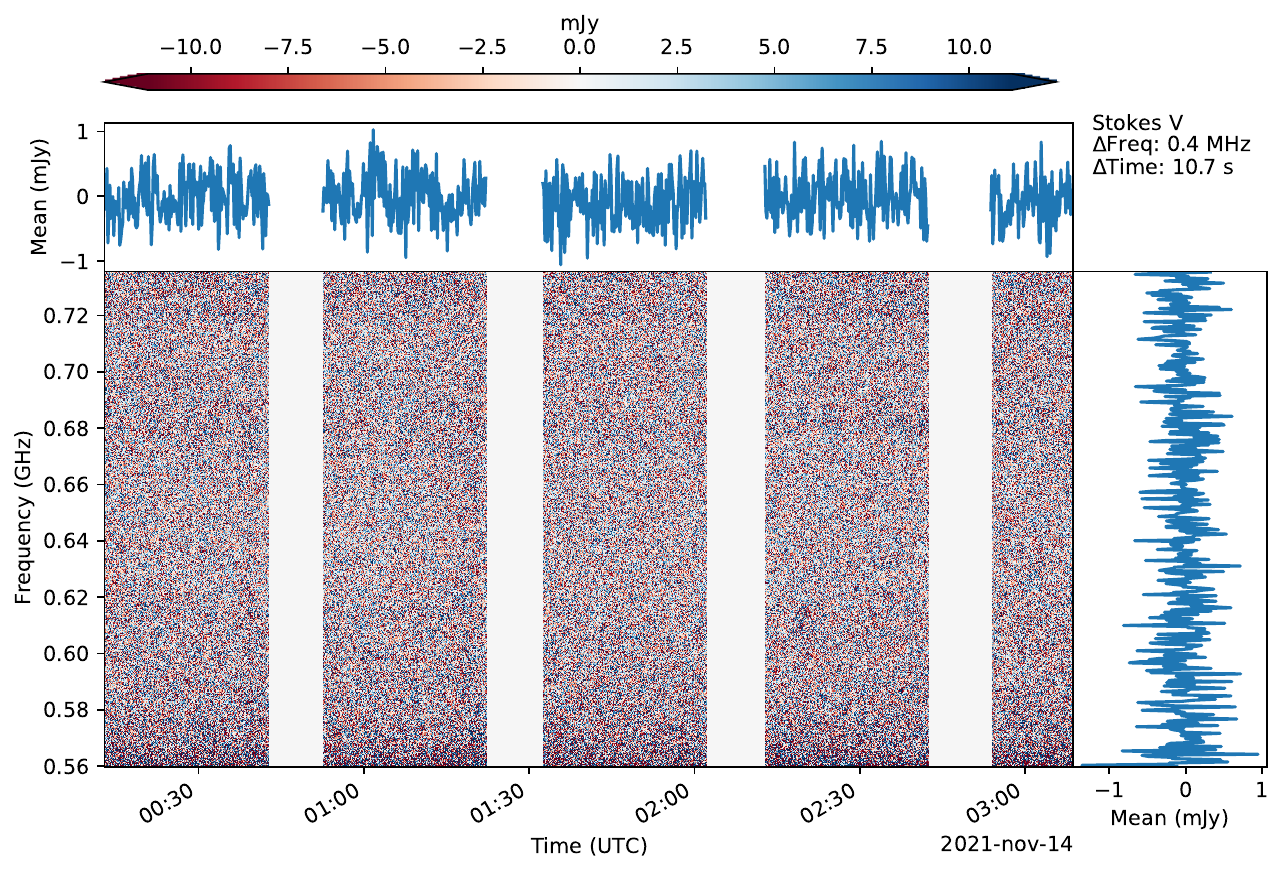}
  \caption{\label{fig:14nov2021_stokes_V_dynspec} \small Stokes V dynamic spectra for 14 November 2021. } 
\label{fig:14nov_dynspec}
\end{subfigure}
\begin{subfigure}{0.45\textwidth}
  \centering
  \includegraphics[width=8cm]{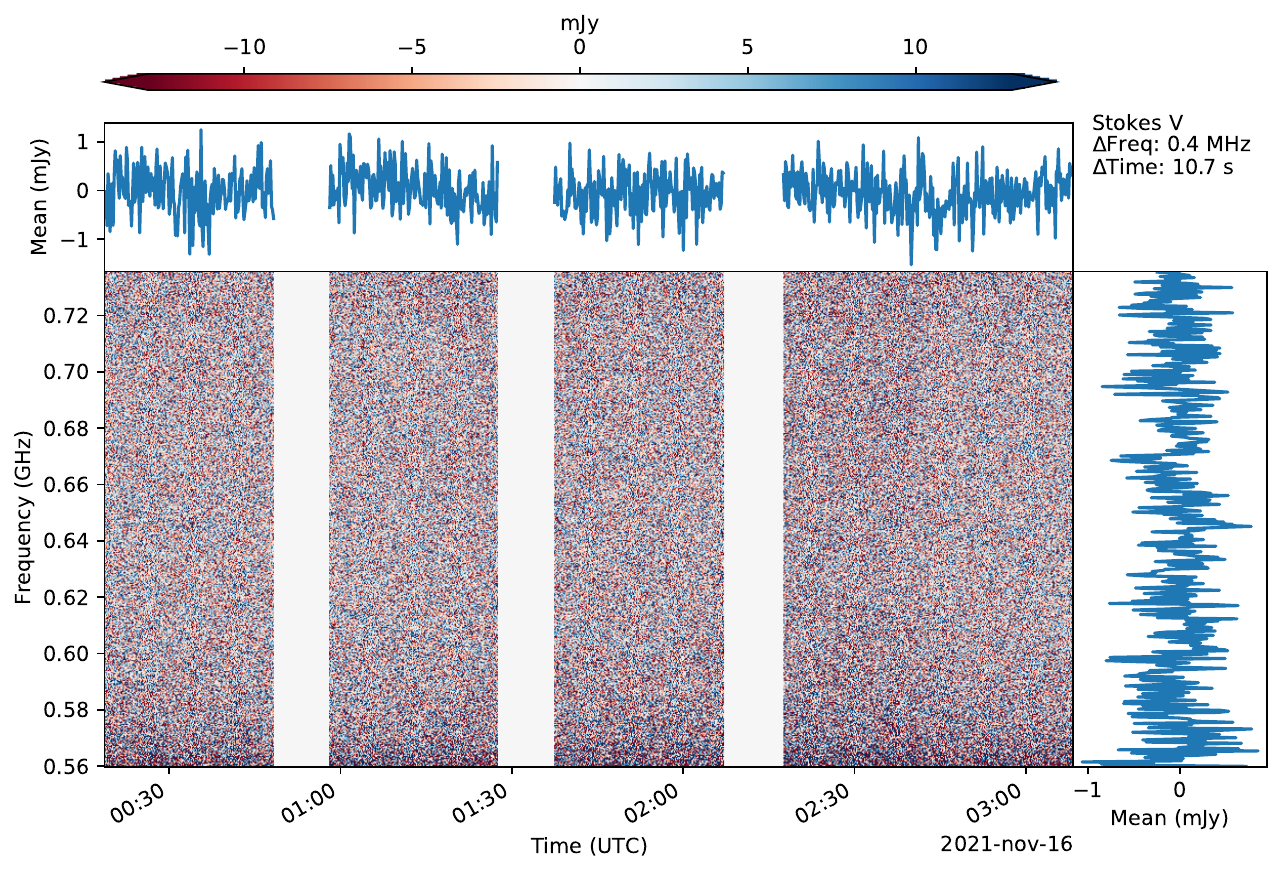}
  \caption{\label{fig:16nov2021_stokes_V_dynspec} \small Stokes V dynamic spectra for 16 November 2021. } 
\label{fig:16nov_dynspec}

\end{subfigure}

\begin{subfigure}{0.45\textwidth}
  \centering
  \includegraphics[width=8cm]{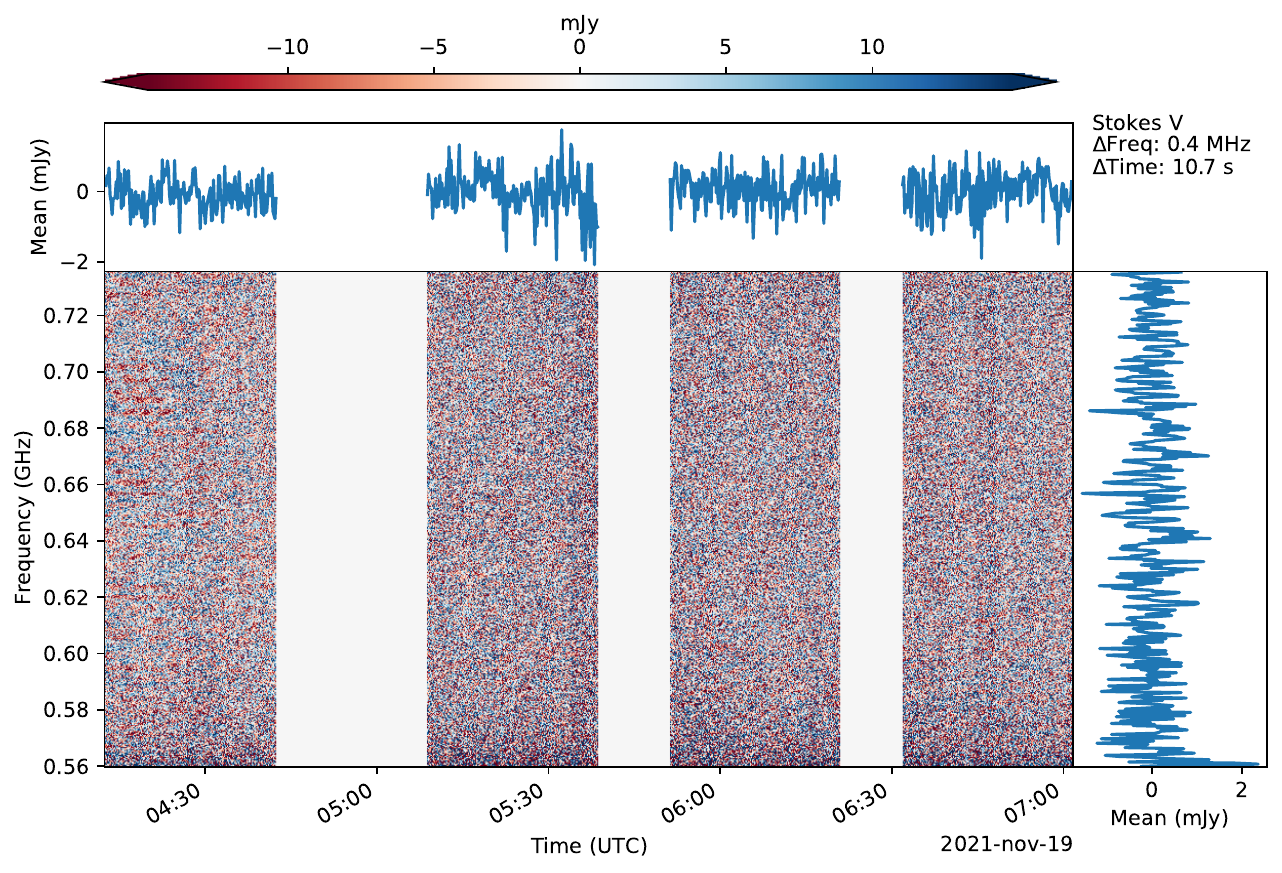}
  \caption{\label{fig:19nov2021_stokes_V_dynspec} \small Stokes V dynamic spectra for 19 November 2021. } 
\label{fig:19nov_dynspec}

\end{subfigure}
\begin{subfigure}{0.45\textwidth}
  \centering
  \includegraphics[width=8cm]{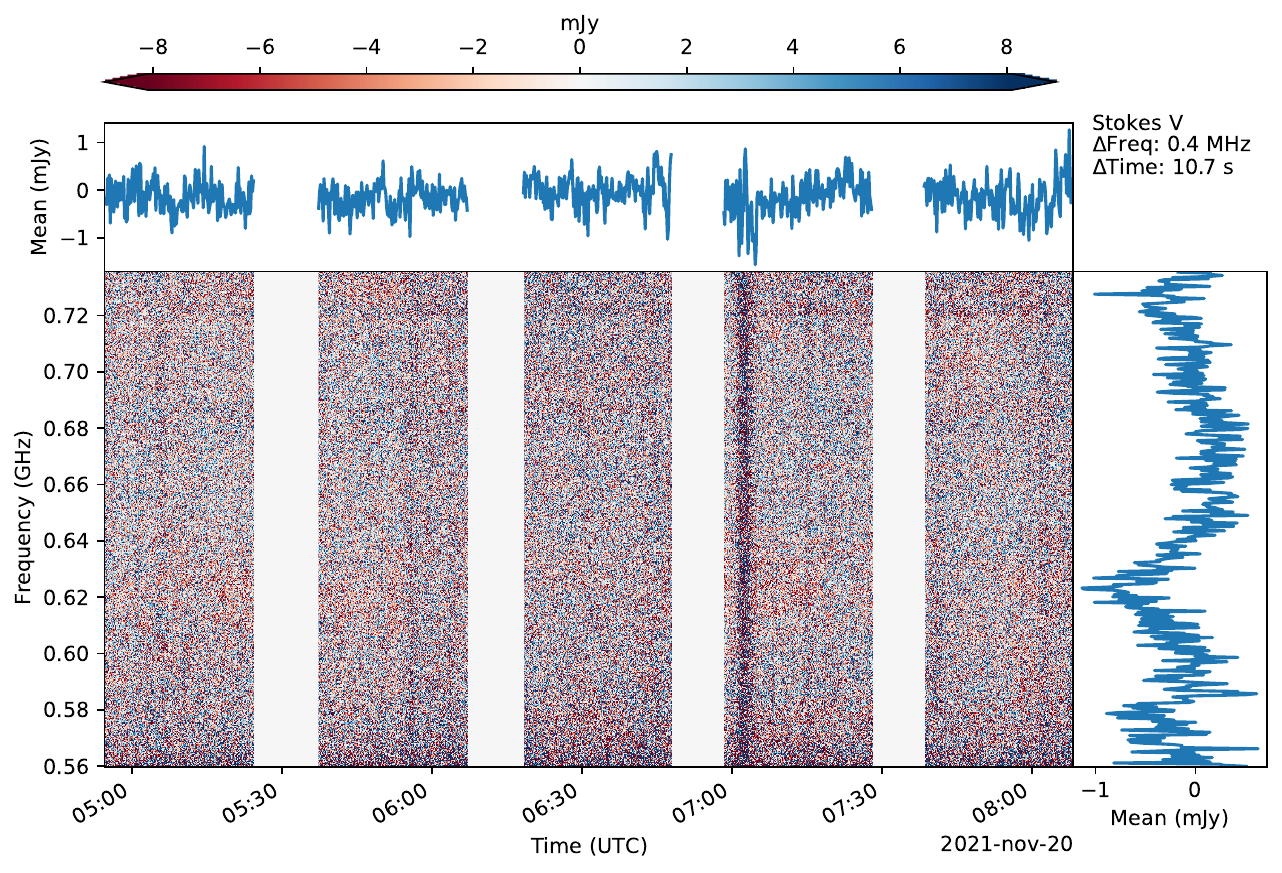}
  \caption{\label{fig:20nov2021_stokes_V_dynspec} \small Stokes V dynamic spectra for 20 November 2021. } 
\label{fig:20nov_dynspec}

\end{subfigure}

\begin{subfigure}{0.45\textwidth}
  \centering
  \includegraphics[width=8cm]{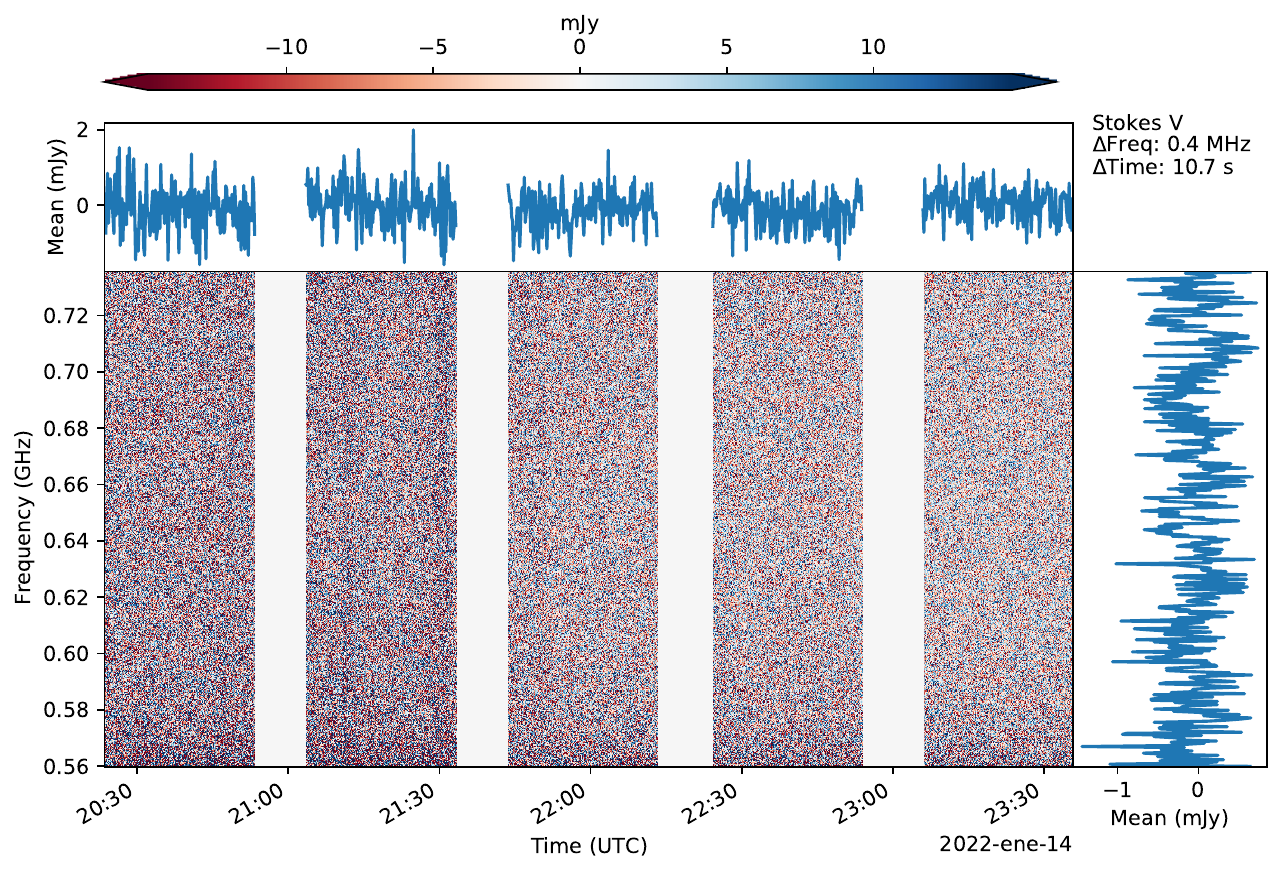}
  \caption{\label{fig:14jan2022_stokes_V_dynspec} \small Stokes V dynamic spectra for 14 January 2022. } 
\label{fig:14jan_dynspec}
\end{subfigure}
\begin{subfigure}{0.45\textwidth}
  \centering
  \includegraphics[width=8cm]{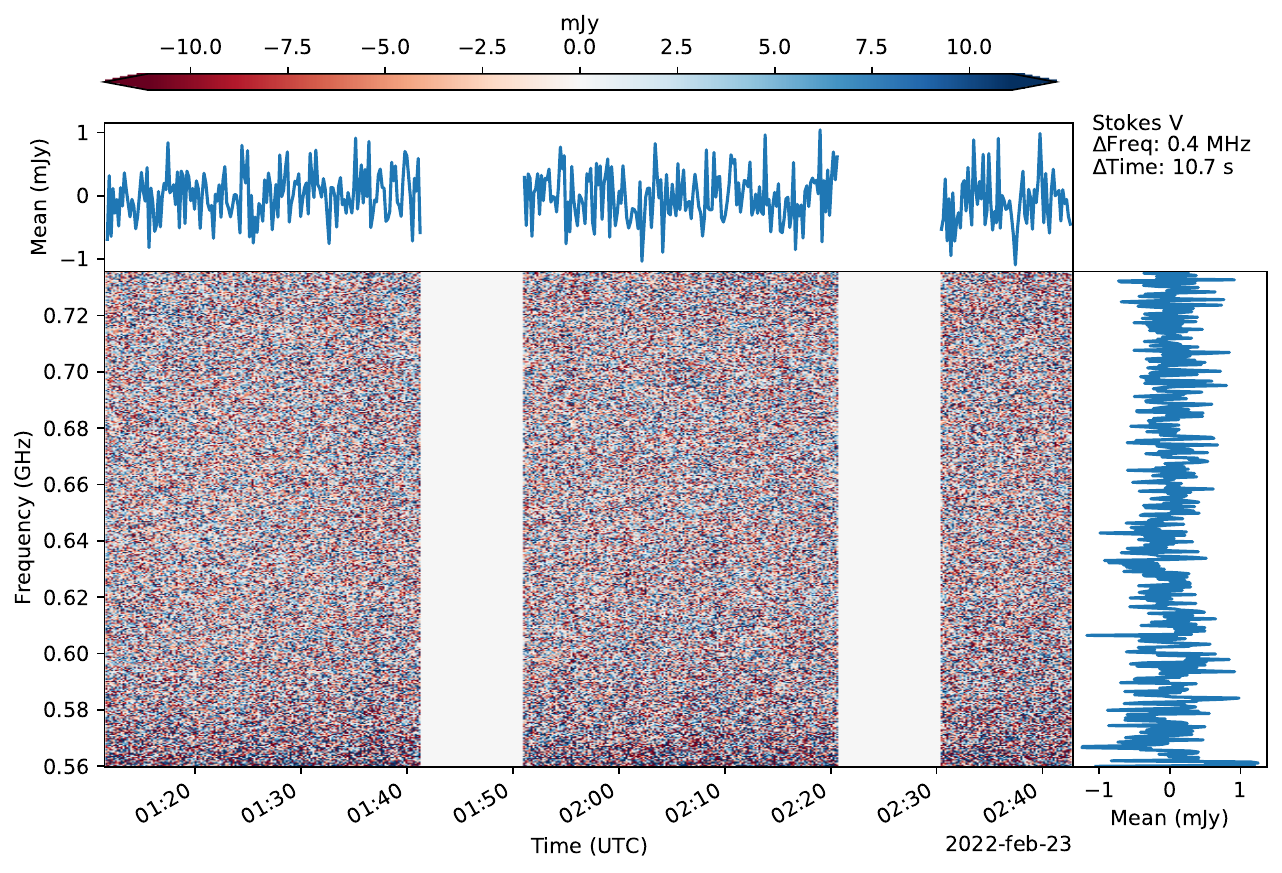}
  \caption{\label{fig:22feb2022_stokes_V_dynspec} \small Stokes V dynamic spectra for 22 February 2022.} 
\label{fig:22feb_dynspec}
\end{subfigure}

\caption{Stokes V dynamic spectra for the rest of the observing sessions. Each dynamic spectra has two other spectra associated, one collapsed in frequency and the other in time.\label{fig:ds_appendix}
}
\label{fig:ds_appendix}
\end{figure*}

\end{appendix}

%%% THE END
\end{document}